\documentclass[aps,prd,superscriptaddress,eqsecnum,nofootinbib,preprintnumbers,reprint,tightenlines,floatfix,titlepage]{revtex4-2}

\usepackage[utf8]{inputenc}
\usepackage{amsmath}
\usepackage{amsfonts}
\usepackage{amssymb}
\usepackage{bm}
\usepackage{CJK}
\usepackage{dsfont}
\usepackage{slashed}
\usepackage[pdftex]{xcolor}
\usepackage{graphicx}
\usepackage{multirow}
\usepackage[sort&compress]{natbib} % To display grouped inline citations e.g. [1-5] instead of [1,2,3,4,5]
\usepackage[normalem]{ulem}

\usepackage[hyperfootnotes=false]{hyperref}
\hypersetup{
    colorlinks=true,     % false: boxed links; true: colored links
    linkcolor=blue,      % color of internal links
    citecolor=blue,      % color of links to bibliography
    filecolor=blue,      % color of file links
    urlcolor=blue,        % color of external links
    linktoc=page
}

\newcommand{\Nlike}{\ensuremath{N}\text{-like}}

\newcommand{\omegalike}{\ensuremath{\Omega}\text{-like}}
\newcommand{\omegaones}{\ensuremath{\Omega_{1}}}
\newcommand{\omegatwos}{\ensuremath{\Omega_{2}}}

\renewcommand{\case}[2]{\ensuremath{{\textstyle\frac{#1}{#2}}}}
\newcommand{\emma}[1]{\ensuremath{\left(#1\right)^\gamma}}
\newcommand{\third}{\ensuremath{{\case{1}{3}}}}

\DeclareMathOperator{\tr}{tr}

\newcommand{\coloaf}{Department of Physics, University of Colorado, Boulder, Colorado 80309, USA}
\newcommand{\fnalaf}{Theory Division, Fermi National Accelerator Laboratory, Batavia, Illinois 60510, USA}
\newcommand{\iuaf}{Department of Physics, Indiana University, Bloomington, Indiana 47405, USA}
\newcommand{\mitaf}{Center for Theoretical Physics, Massachusetts Institute of Technology, \\Cambridge, Massachusetts 02139, USA}
\newcommand{\msuaf}{Department of Computational Mathematics, Science and Engineering, and Department of Physics and Astronomy, Michigan State University, East Lansing, Michigan 48824, USA}

\newcommand{\ugraf}{CAFPE and Departamento de F\'isica Te\'orica y del Cosmos, Universidad de Granada, \\18071 Granada, Spain}
\newcommand{\uiucaf}{Department of Physics, University of Illinois Urbana-Champaign, Urbana, Illinois 61801, USA}
\newcommand{\icasuuiaf}{Illinois Center for Advanced Studies of the Universe, University of Illinois Urbana-Champaign, \\Urbana, Illinois 61801, USA}
\newcommand{\unizar}{Departmento de F\'isica Te\'orica, Universidad de Zaragoza, 50009 Zaragoza, Spain}
\newcommand{\utahaf}{Department of Physics and Astronomy, University of Utah, \\ Salt Lake City, Utah 84112, USA}
\newcommand{\syracuseaf}{Department of Physics, Syracuse University, Syracuse, New York 13244, USA}
\newcommand{\csuaf}{Department of Physics, Colorado State University, Fort Collins, Colorado 80523, USA}
\newcommand{\capa}{Centro de Astropart\'iculas y F\'isica de Altas Energías,  Universidad de Zaragoza, \\50009 Zaragoza, Spain}
\newcommand{\washuaf}{Department of Physics, Washington University, St.~Louis, Missouri 63130, USA}
\newcommand{\apsaf}{American Physical Society, Hauppauge, New York 11788, USA}
\newcommand{\llnlaf}{Nuclear and Chemical Sciences Division, Lawrence Livermore National Laboratory, \\Livermore, California 94550, USA}
\newcommand{\umdaf}{Maryland Center for Fundamental Physics and Department of Physics, \\
University of Maryland, College Park, MD 20742, USA}

\begin{document}
\count\footins = 1000 %stops footnotes overlapping with the page number

\begin{CJK*}{UTF8}{bsmi}
\CJKfamily{bsmi}

\title{High-Precision Scale Setting with the Omega-Baryon Mass and Gradient Flow}

\author{Alexei~Bazavov}\email{bazavov@msu.edu}\affiliation{\msuaf}
\author{Claude~W.~Bernard}\affiliation{\washuaf}
\author{David~A.~Clarke}\affiliation{\utahaf}
\author{Carleton~DeTar}\affiliation{\utahaf}
\author{Aida~X.~\surname{El-Khadra}}\affiliation{\uiucaf}\affiliation{\icasuuiaf}
\author{Elvira~G\'amiz}\affiliation{\ugraf}
\author{Steven~Gottlieb}\affiliation{\iuaf}
\author{Anthony~V.~Grebe}\email{agrebe@umd.edu}\affiliation{\fnalaf}\affiliation{\umdaf}
\author{Urs~M.~Heller}\affiliation{\apsaf}
\author{Leon~Hostetler}\affiliation{\iuaf}
\author{William~I.~Jay}\affiliation{\csuaf}
\author{Hwancheol~Jeong}\affiliation{\iuaf}
\author{Andreas~S.~Kronfeld}\email{ask@fnal.gov}\affiliation{\fnalaf}
\author{Yin~\surname{Lin} (林胤)}\affiliation{\mitaf}
\author{Shaun~Lahert}\affiliation{\utahaf}
\author{Jack~Laiho}\affiliation{\syracuseaf}
\author{Michael~Lynch}\affiliation{\uiucaf}\affiliation{\icasuuiaf}
\author{Andrew~T.~Lytle}\affiliation{\uiucaf}\affiliation{\icasuuiaf}
\author{Aaron~S.~Meyer}\affiliation{\llnlaf}
\author{Ethan~T.~Neil}\affiliation{\coloaf}
\author{Curtis~T.~Peterson}\affiliation{\msuaf}
\author{James~N.~Simone}\affiliation{\fnalaf}
\author{Jacob~W.~Sitison}\affiliation{\coloaf}
\author{Ruth~S.~\surname{Van~de~Water}}\email{ruthv@fnal.gov}\affiliation{\fnalaf}
\author{Alejandro~Vaquero}\affiliation{\utahaf}\affiliation{\unizar}\affiliation{\capa}
\author{Michael~L.~Wagman}\affiliation{\fnalaf}

\collaboration{Fermilab Lattice and MILC Collaborations}
\noaffiliation

\date{\today}

\preprint{FERMILAB-PUB-25-0665-T, LLNL-JRNL-2010752}

\begin{abstract}
The gradient-flow scale $w_0$ in lattice QCD is determined using the mass of the $\Omega^-$ baryon to set the physical scale.
Nine ensembles using the highly improved staggered quark (HISQ) action with lattice spacings of 0.15~fm down to 0.04~fm are used, seven of which have nearly physical light-quark masses.
Electromagnetic corrections to the $\Omega^-$ mass are defined in order to compute a pure-QCD $\Omega$ mass.
The final result is $w_0 = 0.17187(68)$~fm, corresponding to a relative uncertainty of 0.40\% and a central value in good agreement with previous calculations in the literature.
\end{abstract}

\maketitle
\end{CJK*}

\tableofcontents

\section{Introduction}

Quantum chromodynamics (QCD)---the theory of the strong force governing quark and gluon interactions---provides a powerful explanation of the behavior of hadrons over a wide range of energy scales.
At low energy, where the strong coupling $\alpha_s$ becomes large and the theory confines, nonperturbative methods become necessary.
Lattice QCD gives a model-independent method for solving the equations governing strong dynamics by computing numerically the Euclidean path integral on a discretized spacetime grid~\cite{Wilson:1974sk}.

The inputs to a lattice calculation are the bare masses of the quarks $\left\{ m_q \right\}$ and the bare gauge coupling.
The relationship between the bare quark masses and the experimentally accessible hadron masses is not known \emph{a priori}, but the bare quark masses can be tuned to reproduce the hadron spectrum.  Via dimensional transmutation, the bare gauge coupling can be related to the lattice spacing $a$, whose inverse serves as an ultraviolet regulator for the theory.  As a practical matter, since all outputs of a lattice-QCD calculation are dimensionless quantities (e.g., $a$ times a hadron mass), the lattice spacing is needed to compare simulation results with experimental observations.

Scale setting---the determination of the lattice spacing---is therefore essential to most lattice-QCD calculations.  
For percent- and subpercent-level computations, such as the anomalous magnetic moment ($g-2$) of the muon~\cite{Aliberti:2025beg} and Cabibbo-Kobayashi-Maskawa (CKM) matrix elements~\cite{FlavourLatticeAveragingGroupFLAG:2024oxs}, scale setting can contribute substantially to the overall uncertainty budget of the calculation, necessitating a high-precision determination of the scale.
Hadronic contributions to the muon $g-2$ are particularly sensitive to the scale setting due to the quadratic dependence of the integration kernel on the muon mass \cite{DellaMorte:2017dyu}, a dimensionful nonhadronic scale. 

Gradient flow~\cite{Luscher:2010iy}, a procedure for smoothing the gluon field, offers a high-precision and computationally cheap method for relating the lattice spacing to a dimensionful \emph{flow scale} $w_0$, but this flow scale cannot be measured in an experiment and must be further matched to a different experimentally accessible observable.
Therefore, an additional determination of a dimensionful scale from lattice QCD is necessary to relate $w_0$ and thus $a$ to experimental data.
Such a scale must be well known experimentally and also be computable to high precision in lattice QCD.

A commonly used scale is the pion decay constant $f_\pi$ \cite{FermilabLattice:2014tsy,Bazavov:2017lyh}, which is related to the well-measured pion muonic decay rate and suffers from no signal-to-noise problems in lattice calculations.
For subpercent calculations, however, electromagnetic corrections to the pion decay constant become important, requiring information about pion structure beyond the decay constant \cite{DiCarlo:2019thl}.
More seriously, the muonic decay rate yields the product $|V_{ud}|f_\pi$, where the CKM matrix element $|V_{ud}|$ must be obtained from other sources.
Most commonly used is the value obtained from nuclear beta decay~\cite{Hardy:2020qwl}, but it would be unfortunate to rely on nuclear physics for a fundamental ingredient of lattice QCD.

Hadron masses provide other dimensionful scales.
Typically the pseudoscalar meson masses ($M_\pi$, $M_K$,  $M_{D_s}$), which are computed with the highest precision, are used for quark-mass tuning, thus requiring some other choice for determining the scale.
Vector mesons suffer from signal-to-noise problems and are usually resonances, so their masses cannot be determined precisely or unambiguously enough.
Baryon masses provide an alternative.
Nucleon masses have been determined experimentally to more than ten significant digits~\cite{ParticleDataGroup:2024cfk}, but the light quark propagators needed to reconstruct them in lattice QCD are computationally expensive, and the signal-to-noise problem is exacerbated at lighter masses~\cite{Parisi:1983ae, lepage}.
In contrast, the $\Omega^-$ baryon $(sss)$ sits in a mass regime where quark propagators are cheap, making it an attractive candidate for scale setting, and its mass has been measured experimentally to better than 0.02\% precision~\cite{ParticleDataGroup:2024cfk}.
This work presents a lattice-QCD computation of the mass of the $\Omega^-$ baryon at a precision of $0.4\%$, allowing a determination of $w_0$ to the same precision.

This paper is organized as follows.
Section~\ref{sec:background} contains theoretical background on the gradient flow (Sec.~\ref{sec:gradient-flow}) and the construction of baryon operators with staggered quarks (\ref{sec:stagOmega}).
The numerical implementation of the calculations is described in Sec.~\ref{sec:ensembles}.
Section~\ref{sec:analysis} covers all aspects of the data analysis:
computing $w_0/a$ (Sec.~\ref{sec:w0/a}),
correlator fits to obtain $a M_\Omega$ (Sec.~\ref{sec:corrfits}),
the electromagnetic contribution to $M_\Omega$ to obtain a value in pure QCD (Sec.~\ref{sec:pure-QCD-MOmega}),
and the continuum limit of $w_0M_\Omega$ (Sec.~\ref{sec:w0MOmega}).
In Sec.~\ref{sec:conclusions}, we offer some concluding remarks.
The appendices cover Bayesian model averaging (Appendix~\ref{app:BMA}), uncertainty propagation for the retuned $\Omega$ baryon mass (Appendix~\ref{app:stuff}), raw results for $aM_\Omega$, $aM_\pi$, $aM_K$, and $w_0/a$ (Appendix~\ref{app:lattice-results}), and the continuum-limit results for $w_0M_\Omega$ in all flow implementations studied here (Appendix~\ref{app:12w0MOmega}).
The material in Appendices~\ref{app:lattice-results} and~\ref{app:12w0MOmega} should be useful to researchers using the $(2+1+1)$-flavor HISQ ensembles.

\section{Theoretical Background}
\label{sec:background}

\subsection{Gradient flow}
\label{sec:gradient-flow}
Gradient flow is a continuous transformation of the gauge field, introduced in Ref.~\cite{Luscher:2010iy}, that drives the field towards stationary points of the flow action $S_\text{flow}$, \textit{i.e.}, towards smooth solutions that satisfy the classical equations of motion of $S_\text{flow}$.
In this section, we review the theoretical details of gradient flow and its relation to scale setting in lattice QCD. 
In this work, we apply the gradient flow only to the gauge field, so we restrict the discussion accordingly.

\subsubsection{Defining equations}

Consider a gauge-field configuration $A_\mu(x)$ (for now, in the continuum) that serves as an initial condition for the following gradient-flow equation:
\begin{equation}
    \frac{\partial B_\mu}{\partial t}=-\frac{\delta S_\text{flow}}{\delta B_\mu},~~~~~B_\mu(x,0)=A_\mu(x),
\end{equation}
where $t$ is the fictitious flow time and $x$ is a spacetime vector.
If the flow action $S_\text{flow}$ is simply the Yang-Mills action
\begin{equation}
    S_\text{flow}[A] = - \int d^4x \, \frac{1}{2}\tr[G_{\mu\nu}G_{\mu\nu}],
\end{equation}
the flow equation takes the form
\begin{equation}
    \frac{\partial B_\mu}{\partial t}=D_\nu G_{\nu\mu},
\end{equation}
where
\begin{align}
    D_\nu G_{\nu\mu} &= \partial_\nu G_{\nu\mu}+g_0[B_\nu,G_{\nu\mu}],\\
    G_{\nu\mu} &= \partial_\nu B_\mu - \partial_\mu B_\nu + g_0[B_\nu,B_\mu].
\end{align}
Note that the flow time is of mass dimension $-2$.

In lattice gauge theory, the rescaled fields $g_0 A_\mu(x)$ are replaced by the gauge-link variables $U_{x,\mu}$, and the flow equation takes the form
\begin{align}
    \frac{\partial V_{x,\mu}(t)}{\partial t}&=-g_0^2
    \left[\partial_{x,\mu}S_\text{flow}[V]\right]V_{x,\mu}(t), \\
    V_{x,\mu}(0)&=U_{x,\mu},
\end{align}
where $\partial_{x,\mu}S_\text{flow}[V]$ is the appropriate $\text{SU}(3)$-valued derivative with respect to link $V_{x,\mu}$.

With this choice of the flow action, the flow equation is first order in flow time and second order in spacetime variables.
This partial differential equation thus describes a diffusion process that, as flow time proceeds, smooths out short-distance fluctuations of the original gauge field $U_{x,\mu}$.

For the discrete version of the flow action $S_\text{flow}$, this paper employs the Wilson plaquette action or the tree-level Symanzik-improved (plaquette and rectangle) action. The corresponding gradient flow is then referred to as the Wilson or Symanzik flow, respectively.

\subsubsection{Gradient-flow scale \texorpdfstring{$w_0$}{w0}}
\label{sec:w0_theory}

The gradient flow can also provide a reference scale if there is a flow-time-dependent dimensionless quantity that is finite in the continuum limit.
It has been shown in Ref.~\cite{Luscher:2011bx} that the action density evaluated on the flowed gauge field configurations is finite to all orders in $\alpha_s$.
In other words, the flowed action density defines an observable.
As the discretization of the observable does not need to be the same as the one for generating the flow, we denote it $S_\text{obs}[V]$. The appropriate dimensionless quantity is then $t^2\langle S[V]\rangle$ where the averaging is with respect to the gauge-field ensemble.
The reference scale $t_0$ is then defined as a point in flow time where the observable reaches a predefined value~$k$:
\begin{equation}
    t^2\langle S_\text{obs}[V(t)]\rangle\bigg\rvert_{t=t_0} = k.
    \label{eq_cond_t0}
\end{equation}
The most commonly used value for $k$ is the originally proposed $k=0.3$~\cite{Luscher:2010iy}.

The flow of duration $t$ introduces a smearing radius $\sqrt{8t}$. 
Ideally, in order to avoid large cutoff or finite-volume effects, the flow should be used in the regime $a\ll \sqrt{8t}\ll a\,\min(N_s,N_t)$, where $N_s$ ($N_t$) is the number of lattice points in the spatial (temporal) direction. 
The choice $k=0.3$ allows the $t_0$ scale to satisfy this constraint for commonly used lattice spacings and volumes.

At finite lattice spacing, the observable $t^2\langle S_\text{obs}[V(t)]\rangle$ is subject to discretization effects.
Empirical evidence suggests~\cite{BMW:2012hcm} that the slope of $t^2\langle S_\text{obs}[V(t)]\rangle$ is less affected by discretization effects than the quantity itself. A different condition for the reference scale $w_0$ was proposed in Ref.~\cite{BMW:2012hcm}:
\begin{equation}
    W_\text{o}(t) \equiv t\frac{d}{dt}\left[t^2\langle S_\text{obs}[V(t)]\rangle\right]\bigg\rvert_{t=w_0^2} = k
    \label{eq_cond_w0}
\end{equation}
with the same value $k=0.3$.
The subscript ``o'' denotes ``original'' and will be contrasted with a corrected quantity below.

The lattice artifacts affecting the $\sqrt{t_0}$ or $w_0$ scale have three sources: the action $S_\text{MC}$ used in the Markov chain Monte Carlo process to generate the ensemble of gauge-field configurations, the action that generates the flow $S_\text{flow}$, and the action for the observable $S_\text{obs}$.
A perturbative analysis of the discretization effects at tree-level in the finite or infinite volume was carried out in Ref.~\cite{Fodor:2014cpa}. The leading discretization effects in the observable $t^2\langle S_\text{obs}(t)\rangle$ can be represented as an expansion in $a^2/t$:
\begin{equation}
    t^2\langle S_\text{obs}(t)\rangle=\frac{3(N^2-1)g_0^2}{128\pi^2}\left( C(a^2/t) + O(g_0^2)\right),
\end{equation}
where $N$ is the number of colors and
\begin{equation}
    C(a^2/t) = 1 + \sum_{m=1}^\infty C_{2m}\frac{a^{2m}}{t^m}.
\end{equation}
Alternatively, $C(a^2/t)$ can be evaluated numerically, effectively summing all orders in $a^2/t$.
Since this procedure would need to be carried out for every flow time $t$ considered, it is more convenient to use the expansion up to and including the fourth order in $a^2/t$.
The first four coefficients---$C_2$, $C_4$, $C_6$, and $C_8$---have been explicitly computed in the infinite volume limit~\cite{Fodor:2014cpa} for various gauge-flow-observable combinations.

\subsection{\texorpdfstring{$\Omega$}{Omega} baryon with staggered quarks}
\label{sec:stagOmega}

The systematic construction of staggered baryon operators was first introduced in Ref.~\cite{Golterman:1984dn} and later expanded in Ref.~\cite{Bailey:2006zn} to include additional flavor symmetries.
In the following subsections, we provide a high-level summary of the group theory needed to construct lattice operators for $\Omega$.
In the following, we reserve $\Omega^-$ for the physical baryon (with QED included) and $\Omega$ for pure QCD, while introducing further similar notations for staggered baryons that have $\Omega$ properties in the continuum limit.
Interested readers should refer to Refs.~\cite{Golterman:1984dn,Bailey:2006zn} for more details.

\subsubsection{Lattice irreducible representations for \texorpdfstring{$\Omega$}{Omega}}

In this work, we use rooted staggered fermions for sea quarks by taking the fourth root of quark determinants.
The rooting procedure is not applicable for solving quark propagators, so each of the valence quarks still retains four taste copies,
and the simulation is partially quenched. 
In the continuum limit, the four tastes become degenerate and form the $\text{SU}(4)_T$ taste-symmetry group for each quark type. When the taste symmetry is combined with the $\text{SU}(2)_I$ isospin-symmetry group, the internal symmetry of the continuum action becomes $\text{SU}(8)_{ud}\times\text{SU}(4)_s$, where $\text{SU}(8)_{ud}\supset \text{SU}(2)_I \times \text{SU}(4)_T$ is the taste and isospin rotation of up and down quarks and $\text{SU}(4)_s$ is the taste rotation of the strange quark.
To distinguish the staggered tastes from the physical states, we refer to them as \Nlike, \omegalike, etc.~\cite{Lin:2019pia}.

Despite the presence of many superfluous states, operators constructed with all quarks assigned to the same taste must have the same spin and flavor structure as baryons in tasteless QCD in order to satisfy antisymmetrization requirements.
Therefore, observables computed using valence quarks of a single taste must have the same continuum-limit masses as their QCD counterparts~\cite{Bailey:2006zn}.
Because of the enlarged isospin-taste symmetry group of the continuum action, many non-single-taste baryons become degenerate with the single-taste ones in the continuum limit, and, as discussed below, they can then be used to extract properties of the desired baryon.

Operators at finite lattice spacings may be decomposed into irreducible representations (irreps) of the lattice symmetry group and the flavor-symmetry group. For staggered fermions the lattice-symmetry group is called the geometrical timeslice group (GTS)~\cite{Golterman:1984dn}.
The GTS mixes the spacetime symmetry with the taste symmetry and has three fermionic irreps: $8$, $8^\prime$, and $16$, where the number denotes the dimension.
The basis elements of the fermionic representations correspond to the eight corners of equally-spaced three-dimensional unit cubes distributed every other site across an entire spatial timeslice.
The representations also encode additional behavior under rotations, acquiring an additional +1 ($-1$) factor in the case of the 8 ($8'$) representation, or transforming with an additional two-dimensional representation structure in the case of the 16 representation.

The number of \omegalike\ states interpolated by each lattice irrep can be obtained by comparing the subduced continuum single-flavor baryon irrep with the totally symmetric $\text{SU}(4)$ taste representation:
\begin{align}
    \text{SU}(2)_S \times \text{SU}(4)_T &\rightarrow  \text{GTS} , \nonumber\\
    \bigg(\frac{3}{2}, 20_S\bigg) &\rightarrow  2\cdot 8\oplus 2\cdot 8^\prime\oplus 3\cdot 16 ,
    \label{eq:isospin32_sub}
\end{align}
where the notation $2\cdot8'$ means that the $8'$ irrep appears with multiplicity 2.
Numbers are used to denote the dimensions of irreps in all groups except the $\text{SU}(2)_S$ spin group, where the usual spin notation is used.
Subscripts on the irrep dimensions indicate whether the irrep is symmetric ($S$) or mixed symmetry ($M$) as appropriate.
The $20_S$ is the symmetric irrep of $\text{SU}(4)_T$ obtained from the tensor product of three fundamental representations, which contains single-taste \omegalike\ states with physical masses.
The multiplicities obtained from subduction of the continuum-limit symmetry group [Eq.~(\ref{eq:isospin32_sub})] show that the $8$ and $8^\prime$ both interpolate two \omegalike\ multiplets and the $16$ interpolates three.
These \omegalike\ multiplets have degenerate masses in the continuum limit but have taste splittings at finite lattice spacing.

The $8$ and $16$ representations also both interpolate a fictitious $N_s$ baryon multiplet, as seen from the subduction of a totally symmetric spin-$\frac{1}{2}$ representation:
\begin{align}
    \text{SU}(2)_S \times \text{SU}(4)_T &\rightarrow \text{GTS} , \nonumber\\
    \bigg(\frac{1}{2}, 20_M\bigg) &\rightarrow 3\cdot 8\oplus 0\cdot 8^\prime\oplus 1\cdot 16.
\label{eq:isospin12_sub}
\end{align}
The $N_{s}$ is a spin-$\frac{1}{2}$ \Nlike\ state formed from two distinct flavors both of mass equal to that of the strange quark.
It has a lower mass than the \omegalike\ states.
The $N_{s}$ states are only possible because the mixed-symmetric taste symmetry (in the $20_M$ representation) allows for a mixed-symmetric spin representation (in the spin-$\frac{1}{2}$ representation) that satisfies overall wavefunction symmetrization.
Without taste symmetry, the spin-$\frac{1}{2}$ representation would vanish due to the overall wavefunction antisymmetrization required by fermion statistics, which is why these states are not present in tasteless QCD.
The existence of the $N_{s}$ state motivates our choice of analysis strategy: we use only the $8^\prime$ irrep in this work since it does not interpolate $N_s$.\footnote{See Refs.~\cite{Meyer:2016kwb,Hughes:2019ico} for information on the $8$ and $16$ irreps.}

\begin{widetext}
The relevant subduction for the \omegalike\ states with $\text{SU}(3)$ flavor symmetry is
\begin{align}
    \text{SU}(2)_S \times \text{SU}(12)_{FT} &\rightarrow \text{SU}(3)_F \times \text{GTS} , \nonumber\\
    \bigg(\frac{3}{2}, 364_S\bigg) &\rightarrow 2\cdot(10_S, 8) \oplus 2\cdot(10_S, 8^\prime) \oplus 3\cdot(10_S, 16) \oplus (8_M, 8^\prime) \oplus \cdots ,
    \label{eq:su3f_subduction}
\end{align}
\end{widetext}
where $\text{SU}(12)_{FT}$ denotes the flavor-taste-symmetry group, and $364_{S}$ is the symmetric irrep thereof that contains the single-taste \omegalike\ states with three strange quarks of identical flavor and taste,
so all states within the irrep are degenerate with the physical $\Omega$ baryon in the continuum and $\text{SU}(3)$-flavor-symmetric limits.
On the right side of the equation, the first three irreps correspond to those obtained in Eq.~\eqref{eq:isospin32_sub}.
The last irrep, $(8_M, 8^\prime)$, is another consequence of the additional taste symmetry with no corresponding representation in tasteless QCD.
This irrep interpolates a single \omegalike\ multiplets and no $N_{s}$ multiplets.
Other omitted irreps, indicated by the ellipsis in Eq.~\eqref{eq:su3f_subduction}, interpolate both $N_s$ and \omegalike\ states and are not used in this work.

We can construct more interpolators of \omegalike\ states by taking advantage of the freedom to adjust quark masses in partially quenched simulations.
In particular, we can use the operators in the irrep $(8_{M},8^\prime)$ of Eq.~(\ref{eq:su3f_subduction}) with all three quark masses set to the physical strange-quark mass.
This construction requires at least two flavors of valence strange quarks to satisfy antisymmetrization, which leads to a valence-sector symmetry group identical in construction to the usual isospin symmetry with light quarks.
We refer to this valence-sector symmetry containing two valence strange quarks as strange isospin to distinguish it from the isospin of light quarks and denote the corresponding flavor-symmetry group by $\text{SU}(2)_{I_{s}}$.
We distinguish the \omegalike\ states produced from strange-isospin interpolating operators by denoting them \omegatwos\ states, in contrast to \omegaones\ states produced from a valence sector containing only a single strange-quark flavor.

The subduction of the two-strange-valence-flavor theory produces the mapping
\begin{align}
    \text{SU}(3)_F &\rightarrow \text{SU}(2)_{I_{s}} , \nonumber\\
    10_{S} &\to 4_{S} \oplus \cdots \\
    8_{M} &\to 2_{M} \oplus \cdots, \nonumber
\end{align}
where only irreps containing interpolating operators constructed exclusively from the two valence strange quarks are listed.
The continuum symmetry group of the strange-isospin valence sector, including the spin group, then becomes $\text{SU}(2)_S\times \text{SU}(8)_{I_{s}T}$, where $\text{SU}(8)_{I_{s}T} \supset \text{SU}(2)_{I_{s}} \times \text{SU}(4)_T$ is the strange-isospin-taste group.
The relevant subduction for the \omegatwos\ states is
\begin{widetext}
\begin{align}
    \text{SU}(2)_S \times \text{SU}(8)_{I_{s}T} &\rightarrow \text{SU}(2)_{I_{s}} \times \text{GTS} , \nonumber\\
    \bigg(\frac{3}{2}, 120_S\bigg) &\rightarrow 2\cdot(4_S, 8) \oplus 2\cdot(4_S, 8^\prime) \oplus 3\cdot(4_S, 16) \oplus (2_M, 8^\prime) \oplus \cdots ,
    \label{eq:su2is_subduction}
\end{align}
\end{widetext}
which are in a one-to-one mapping with the irreps in Eq.~(\ref{eq:su3f_subduction}).
The last irrep, $(2_{M},8^\prime)$,
 is the only strange-isospin interpolating operator used in this work.
The relevant subduction for the \omegaones\ states, which are a subset of the interpolating operators in the $(4_{S},8^\prime)$ irrep of Eq.~(\ref{eq:su2is_subduction}), are the same as those in the $8^\prime$ irrep of Eq.~(\ref{eq:isospin32_sub}).

In summary, by subducing irreps of the continuum-symmetry group down to irreps of the lattice-symmetry group, we arrive at two irreps that will be used in this work to compute the $\Omega$ mass. These are
\begin{enumerate}
    \item the flavor-symmetric $8^\prime$ irrep that interpolates two taste-split \omegaones\ multiplets as the lowest-energy multiplets,
    \item the strange-isospin $8^\prime$ mixed-symmetry irrep that interpolates to one \omegatwos\ multiplets as the ground state.
\end{enumerate}
Explicit expressions for operators transforming in these irreps are presented in Sec.~\ref{sec:omega-baryon-correlators} below.
In the continuum limit, all three \omegalike\ multiplets should have the same masses.
Other operators that interpolate to \omegalike\ states have more complicated spectra including lower-lying $N_{s}$ states, and we do not consider them here.

\section{Numerical Implementation}
\label{sec:ensembles}

\subsection{Gauge-field ensembles}

The calculations described in this work were carried out using nine $(2+1+1)$-flavor ensembles generated primarily by the MILC collaboration~\cite{MILC:2012znn,Bazavov:2017lyh}, listed in Table~\ref{tab:ensembles-used}.
One of the ensembles, 0.09C, was generated by the combined efforts of the CalLat~\cite{Miller:2020evg}, MILC, and PNDME collaborations.
The highly improved staggered-quark (HISQ) action~\cite{Follana:2006rc} was used for both sea and valence quarks.
The sea-quark masses for these ensembles have been tuned to reproduce the physical $\pi$-, $K$-, and $D_s$-meson masses to within few-percent accuracy (except for 0.12H and 0.09H, which have a heavier-than-physical $am_l^\text{sea}$ to study sea-quark mistuning), and the lattice spacings span from 0.15~fm down to about 0.04~fm.

\begin{table*} 
    \centering
    \newcommand{\h}{\phantom{H}}
    \caption{List of ensembles used in this work.
    The three near-physical ensembles at $a \approx 0.09$~fm are the original, slightly mistuned (M) ensemble; a corrected (C) ensemble with closer-to-physical meson masses; and a large-volume (L) ensemble designed to study finite-volume effects.
    Ensembles labeled with H represent a heavier-than-physical pion mass of about 220~MeV and were used to study pion-mass mistuning.
    On four of the ensembles, note that two valence strange quark masses were used for studying the effects of quark-mass mistuning.
    Values for meson masses on the physical-point ensembles (0.15, 0.12, 0.09M, 0.09C, 0.06, and 0.04~fm) are taken from Ref.~\cite{MILC:2024ryz} and references therein.
    Values on the heavier-than-physical ensembles 0.12H and 0.09H are taken from Refs.~\cite{FermilabLattice:2022gku} and~\cite{Bazavov:2017lyh}, respectively.
    The value on the large-volume ensemble 0.09L is new.
    In all cases, values are determined from multi-exponential fits to the truncated spectral decomposition using methods and parameters described in Refs.~\cite{Bazavov:2017lyh,FermilabLattice:2022gku}.
    The number of configurations, total number of sources, and number of independent samples after binning are also shown.}
    \label{tab:ensembles-used}
    \begin{tabular}{cccccccccccc}
        \hline\hline
        $\approx a$/fm & $\approx L$/fm & $N_s^3 \times N_t$ & $am_l^\text{sea}$ & $am_s^\text{sea}$ & $am_c^\text{sea}$ & $am_s^\text{val}$ & $am_\pi$ & $am_K$ & $N_\text{cfg}$ & $N_\text{src}/10^3$ &  $N_{\rm bin}$  \\
        \hline
        0.15\h &  4.8 &  $32^3 \times 48$  & 0.002426 & 0.0673 & 0.8447 & $\begin{matrix} 0.0673 \\ 0.067973 \end{matrix}$ & 0.103414(11) & 0.37847(11) & 9936 & 238 & {331 }\\
        0.12H  &  4.8 &  $40^3 \times 64$  & 0.00507 & 0.0507 & 0.628 & 0.0507 & 0.13426(8) & 0.30806(12) & 1030 & 33 &  {51 } \\ 
        0.12\h &  5.8 &  $48^3 \times 64$  & 0.001907 & 0.05252 & 0.6382 & $\begin{matrix} 0.05252 \\ 0.0530452 \end{matrix}$ & 0.0830651(63) & 0.303949(77) & 9995 & 320 & {333 } \\
        0.09H  &  4.3 &  $48^3 \times 96$  & 0.00363 & 0.0363 & 0.432 & 0.0363 & 0.09860(13) & 0.22706(15) & 979 & 94 & {48 } \\
        0.09M  &  5.8 &  $64^3 \times 96$  & 0.00120 & 0.0363 & 0.432 & $\begin{matrix} 0.0363 \\ 0.03636 \end{matrix}$ & 0.057184(30) & 0.219482(70) & 5333 & 256 & {266 }\\
        0.09C  &  5.8 &  $64^3 \times 96$  & 0.001326 & 0.03636 & 0.4313 & $\begin{matrix} 0.03636 \\ 0.0367236 \end{matrix}$ & 0.060069(26) & 0.220231(40) & 4692 & 196 & {234 }\\
        0.09L  & 11.5 & $128^3 \times 96$  & 0.001326 & 0.03636 & 0.4313 & 0.03636 & 0.0600523(72) & 0.220225(13) & 1051 & 50 & {52 }\\
        0.06\h &  5.8 &  $96^3 \times 192$ & 0.0008 & 0.022 & 0.260 & 0.022 & 0.038842(29) & 0.142607(51) & 4061 & 390 & {135 }\\
        0.04\h &  5.8 & $144^3 \times 288$ & 0.000569 & 0.01555 & 0.1827 & 0.01555 & 0.028981(18) & 0.106297(39) & 1007 & 145 & {50 }\\
        \hline\hline
    \end{tabular}
\end{table*}

\subsection{Gradient-flow observables}
In this work, we consider Wilson (plaquette), Symanzik (plaquette and rectangle), and clover discretizations of the observable $S_\text{obs}$. Together with the two flows it gives us six combinations that are labeled as gauge-flow-observable: SWW, SWS, SWC, SSW, SSS, SSC. The first letter, S, is the same as it corresponds to the action used to generate the MILC ensembles that are used in this study: the one-loop Symanzik-improved gauge action~\cite{Symanzik:1979ph,Symanzik:1983dc,Weisz:1982zw,Weisz:1983bn,Curci:1983an,Luscher:1984xn,Luscher:1985zq} and the highly improved staggered quark (HISQ) action~\cite{Follana:2006rc,Hao:2007iz,Hart:2008sq}. 
The second and third letters denote the actions used for flowing the gauge field and measuring the observable, respectively.
The tree-level corrections of Ref.~\cite{Fodor:2014cpa} can be applied to the flowed observable $S_\text{obs}$ to cancel partially the discretization effects. This leads to the following corrected scale conditions:
\begin{align}
    \left.\frac{t^2\langle S_\text{obs}(t)\rangle}{1+\sum_{m=1}^4C_{2m}(a^2/t)^m}\right|_{t=t_0} &= k,
    \label{eq_cor_cond_t0} \\
    W_\text{c}(t) \equiv \left.t\frac{d}{dt}\frac{t^2\langle S_\text{obs}(t)\rangle}{1+\sum_{m=1}^4C_{2m}(a^2/t)^m}\right|_{t=w_0^2} &= k.
    \label{eq_cor_cond_w0}
\end{align}
For the scales determined from the original definitions (\ref{eq_cond_t0}) and (\ref{eq_cond_w0}), the letter ``o'' (original) is added to the gauge-flow-observable combinations; for scales determined from the tree-level corrected definitions (\ref{eq_cor_cond_t0}) and (\ref{eq_cor_cond_w0}), the letter ``c'' (corrected) is added. 
The nomenclature for the scales determined from various gauge-flow-observable discretizations with or without the tree-level corrections applied is then S[W,S][W,S,C][o,c].
Previous work from MILC on the scales $\sqrt{t_0}$ and $w_0$~\cite{MILC:2015tqx} included only the SSCo and SSCc combinations.
\pagebreak

To integrate the gradient flow, we use a fourth-order integrator described in Refs.~\cite{Bazavov2022LieInt,Bazavov:2021pik} with the coefficient scheme of Ref.~\cite{BERLAND20061459}.
To control the systematic step-size errors in the flow integration, we perform all measurements with two step sizes $\Delta t=1/20$ and $\Delta t = 1/40$.
The integration is stable on all ensembles and the systematic errors coming from the flow integration are always below the statistical errors by at least two orders of magnitude.

\subsection{\texorpdfstring{\boldmath$\Omega$}{Omega}-baryon correlators\label{sec:omega-baryon-correlators}}

Using the notation of Refs.~\cite{Golterman:1984dn,Bailey:2006zn}, the basic building block of staggered baryon operators with zero momenta is the quark trilinear
\begin{align}
    _{ijk}B_{A,B,C} &\equiv \sum_{x_l\in\text{even}} \frac{1}{6}\epsilon_{abc}D_A \chi^a_i(x) D_B\chi^b_j(x) D_C \chi^c_k(x),\\
    D_A\chi^a_i(x) &\equiv \frac{1}{2}\big(
    \chi^a_i(x-\hat{A}) + \chi^a_i(x+\hat{A})
    \big).
    \label{eq:shifted-op}
\end{align}
$a,b,c$ are color indices, $i,j,k \in \{s, s^\prime\}$ are the strange-isospin indices for two valence strange quarks $s$ and $s^\prime$, and $A, B, C \in \{0, 1, 2, 3, 12, 13, 23, 123\}$ denote one of the eight corners of a three-dimensional spatial cube with side length $a$ and $\hat{A}$ is the corresponding shift vector.
To get a zero-momentum state, we sum over all the even spatial indices as dictated by the symmetry of the staggered action.
Because the shifts in Eq.~\eqref{eq:shifted-op} are not accompanied by gauge links, the quark trilinear is not gauge invariant; suitable gauge fixing is described below.

For flavor-symmetric $8^\prime$ irreps coupling to \omegaones\ states, $i=j=k=s$ and there are two different classes of operators~\cite{Golterman:1984dn} belonging to the same lattice irrep.
These are 
\begin{align}
    \mathcal{S}_{4} &\equiv {}_{sss}B_{1,12,13} - {}_{sss}B_{2,21,23} + {}_{sss}B_{3,31,32},\\
    \mathcal{S}_{7} &\equiv {}_{sss}B_{1,2,3},
\end{align}
where $\mathcal{S}$ denotes that the irreps are symmetric and the subscript denotes the class of the operators as enumerated in Ref.~\cite{Bailey:2006zn}.
On the other hand, there is a single class-$4$, strange-isospin, $8^\prime$ operator coupling to the \omegatwos\ state formed from the mixed-symmetric combination of flavor indices:
\begin{widetext}
\begin{align}
    \mathcal{M}_4 &\equiv 
    {}_{sss'}M_{1,12,13} - 
    {}_{sss'}M_{2,21,23} +
    {}_{sss'}M_{3,31,32} +
    {}_{ss's}M_{1,12,13} -
    {}_{ss's}M_{2,21,23} +
    {}_{ss's}M_{3,31,32},
    \\
    _{ijk}M_{A,B,C} &\equiv 
    \frac{1}{3}\big(
    2 \,{}_{ijk}B_{ABC} - {}_{jki}B_{ABC} + {}_{kij}B_{ABC}
    \big).
\end{align}
\end{widetext}

Reference~\cite{Borsanyi:2020mff} uses the same set of \omegalike\ operators for computing the $\Omega$ baryon mass.
In particular, their $\Omega_\text{VI}$, $\Omega_\text{XI}$, and $\Omega_\text{Ba}$ operators are the same as our $\mathcal{S}_4$, $\mathcal{S}_7$, and $\mathcal{M}_4$ operators, respectively, up to normalization factors.

The quark propagators in this work were computed between a corner wall source and a point sink.
The corner wall source is defined to be 1 on one of the eight corners of each of the three-dimensional spatial cubes on a timeslice (specified by $A,B,C$ above) and zero elsewhere.
Since none of the $\Omega$-baryon interpolating operators described in the previous section make use of the $0$ or $123$ corners of the hypercube, only six quark propagators must be computed for each source timeslice.
The various shifts required by Eq.~(\ref{eq:shifted-op}) are included at the sink.
Since the operations at both source and sink are not gauge invariant, the configurations were fixed to Coulomb gauge.

For the four most computationally intensive ensembles---0.09M, 0.09L, 0.06, and 0.04~fm---the truncated solver method~\cite{Collins:2007mh,Bali:2009hu} was used to reduce the computational cost of propagator solves.
Sloppy solves were performed with precision $\varepsilon_\text{sloppy} = 10^{-4}$, and a single fine propagator solve with a precision of $\varepsilon_\text{fine} = 10^{-12}$ was performed on each configuration for bias correction.
For the coarser ensembles, where quark propagators are computationally cheap, all solves were performed at the fine precision.
The total number of sources used for each ensemble is given in Table~\ref{tab:ensembles-used}.

\subsection{Electromagnetism}
\label{sec:em-setup}

Although the electromagnetic contribution to the $\Omega^-$ baryon mass is expected to be small, the precision sought here requires care.
Our approach follows the electroquenched approximation~\cite{Duncan:1996xy} in which the charges of the sea quarks are neglected and the valence quarks interact with free photon fields.
The MILC Collaboration has previously used this approach to compute electromagnetic effects for mesons~\cite{MILC:2015ypt,MILC:2018ddw}. 
For the Abelian U(1) group, we choose the noncompact formulation.

Thus, the electromagnetic gauge action is
\begin{equation}
    S_\text{em} = \frac{1}{4} \sum_{n,\mu,\nu} \left[ \partial_\mu A_\nu(n) - \partial_\nu A_\mu(n)\right]^2.
\end{equation}
with Coulomb gauge-fixing $\partial_i A_i(n)=0$.
The momentum-space distribution is Gaussian, so the $A_\mu(k)$ can be independently and trivially generated (with no need for a Markov chain) and converted to position space via a fast Fourier transform.
To obtain a convergent path integral, it is necessary to drop zero modes.
We choose the $\text{QED}_{TL}$ formulation~\cite{Duncan:1996xy,BMW:2014pzb}, namely,
\begin{align}
    A_0 (k_0,\bm{k}=\bm{0}) &= 0, \qquad \forall k_0 \nonumber \\
    \bm{A}(k_0=0,\bm{k}=\bm{0}) &= \bm{0} , \\
    \bm{k} \cdot \bm{A}(k_0,\bm{k}) &= 0, \qquad \forall k_0,\bm{k} , \nonumber
\end{align}
with the last condition enforcing Coulomb gauge.
More details can be found in Ref.~\cite{MILC:2018ddw}.

The Coulomb-gauge $A_\mu$ field is promoted to a U(1) link variable $U_{\mu}^\text{em}(n) = e^{ieqA_\mu(n)}$, with $q$ the electric charge of the quark in units of the positron charge~$e$.
The gauge-fixed SU(3) links are multiplied by the U(1) link phases before HISQ smearing is carried out and the solver is called.
A different U(1) field is generated for each source timeslice used.

\section{Data Analysis}
\label{sec:analysis}

\subsection{Computing \texorpdfstring{\boldmath$w_0/a$}{w0/a}}
\label{sec:w0/a}

To extract $w_0/a$ on each ensemble, we compute the ensemble average and the uncertainty for the quantity $W(t)$ on the left side of Eq.~(\ref{eq_cond_w0}) or Eq.~(\ref{eq_cor_cond_w0}) at all flow times (which are on a grid with spacing $\Delta t$).
The derivative with respect to $t$ is approximated with a finite difference computed using a five-point stencil that is valid up to order~$(\Delta t)^4$.
On the grid of points where $W(t)$ is computed, let $\tilde t$ be the one where $W(t)$ is closest to the number~$k$ appearing on the right-hand side of  Eq.~(\ref{eq_cond_w0}) or Eq.~(\ref{eq_cor_cond_w0}).
We then fit $W(t)$ in the range $[\tilde t - n\Delta t, \tilde t + n\Delta t]$ with a fourth order polynomial $p_4(t)$, where the number $n$ ranges from 3 to 6 depending on the ensemble.
Solving $p_4(t)=k$ for $t=(w_0/a)^2$ (which generically lies in between the original grid points) then provides $(w_0/a)^2$ to fourth-order accuracy in~$\Delta t$.

To propagate the statistical uncertainty on $w_0/a$ we use a jackknife binning procedure.
We use 10--200 jackknife bins, depending on the length of the ensemble, and extrapolate the uncertainty to infinite bin size. We also directly measure the integrated autocorrelation time for the action density $S_\text{obs}$ at the flow time close to $(w_0/a)^2$ on various ensembles, as has been reported earlier~\cite{Bazavov:2024dov}. We observe that the extrapolation to infinite bin size provides the most conservative uncertainty estimate, and we report it as the final statistical uncertainty on $w_0/a$ in Table~\ref{tab:w0-scale}.

\subsection{Correlator fits}
\label{sec:corrfits}

The correlation functions $C(t)$ for the $\Omega$ baryon at source-sink separation\footnote{In this subsection, $t$ refers to distances in the Euclidean time direction of the space-time lattice,  rather than the gradient flow time underlying the observables of Eqs. (\ref{eq_cor_cond_t0})-(\ref{eq_cor_cond_w0}). } $t$ are fit to the spectral decomposition in the form
\begin{equation}
    f(t) = \sum_{i=1}^{N_+} c_i^{(+)} e^{-E_i^{(+)} t} + \sum_{i=1}^{N_-} c_i^{(-)} e^{-E_i^{(-)} t} (-1)^t ,
    \label{eq:fit-function}
\end{equation}
which has $N_+$ decaying (nonoscillating) states and $N_-$ oscillating states, which correspond to negative-parity staggered baryons.  In this analysis, all fits considered have equal numbers of oscillating and nonoscillating states, that is, $N_+ = N_-$.  We define a $\chi^2$ function
\begin{equation}
    \chi^2 = \sum_{tt'} [f(t) - C(t)] \Sigma^{-1}_{tt'}[f(t') - C(t')] ,
    \label{eq:chi2}
\end{equation}
where $C(t)$ denotes correlator data; $\Sigma$ is the corresponding sample covariance matrix; and $t$, $t'$ range over the timeslices fitted.

Correlation functions at various source times are first averaged on each gauge configuration.  Significant autocorrelations are observed on the source-averaged correlators, corresponding to an integrated autocorrelation time $\tau_\text{int}$ of about 5--10 configurations.  To correct for this, correlators are binned by 20--25 configurations depending on the ensemble, after which autocorrelations are negligible in the binned stream.  The statistical uncertainty in the binned correlators are observed to be stable with additional binning.

However, some ensembles contain only about 1000 configurations, so only $N_\textrm{bin} \approx 50$ independent samples remain after binning by 20 (as shown in Table~\ref{tab:ensembles-used}), an insufficient number to estimate the covariance matrix $\Sigma$ to a sufficient precision to compute its inverse reliably.  This can be ameliorated by thinning the data to limit the number of timeslices $N_\text{fit}$ in the fit, thereby ensuring $N_\text{fit}^2/2 \lesssim N_\text{bin}$.  Due to the oscillating states in Eq.~(\ref{eq:fit-function}), the fits become unstable if every other timeslice is removed (as the oscillating and nonoscillating states become impossible to distinguish).
However, removing every third timeslice or two out of every three timeslices preserves the oscillating structure while decreasing $N_\text{fit}$ below $\sim \sqrt{2 N_\text{bin}}$.

Priors are added to the fit by augmenting the $\chi^2$ function in Eq.~(\ref{eq:chi2}) with Bayesian priors,
\begin{equation}
    \chi^2_\text{aug} = \chi^2_\text{data} + \sum_i \frac{\left(c_i - c_i^{(0)}\right)^2}{(\Delta c_i)^2} + \sum_i \frac{\left(E_i - E_i^{(0)}\right)^2}{(\Delta E_i)^2} ,
\end{equation}
where $c_i^{(0)}$ and $E_i^{(0)}$ are prior central values, and $\Delta c_i$ and $\Delta E_i$ are prior widths.
In the \verb|corrfitter| fitting framework \cite{gvar,lsqfit,corrfitter}, these priors are implemented effectively as additional data points; that is, the covariance matrix $\Sigma$ is augmented to include the prior contribution as well.  For numerical stability, an SVD cut of $10^{-12}$ is imposed on the covariance matrix of each ensemble.

The priors are chosen to improve the numerical stability of the fit without biasing the $\Omega$-baryon fit results.  Thus, the prior widths are chosen to be much wider than the posterior widths.
The central value for the ground-state energy prior is chosen based on the time-averaged effective mass in the plateau region, where time averaging damps the contribution of the oscillating states \cite{Bailey:2008wp}.
Priors for gaps to excited-state energies are estimated from positive- and negative-parity excitations of the $\Omega$ baryon in the Particle Data Group (PDG) tables~\cite{ParticleDataGroup:2024cfk} similarly to Ref.~\cite{Borsanyi:2020mff}.
The overlap factors in the spectral decomposition (i.e., the coefficients of the exponentials in Eq.~(\ref{eq:fit-function}))---nuisance parameters from the standpoint of energy determination---are given very wide priors with widths exceeding 100\% of the central value.
\begin{figure}
    \centering
    \includegraphics[width=\linewidth]{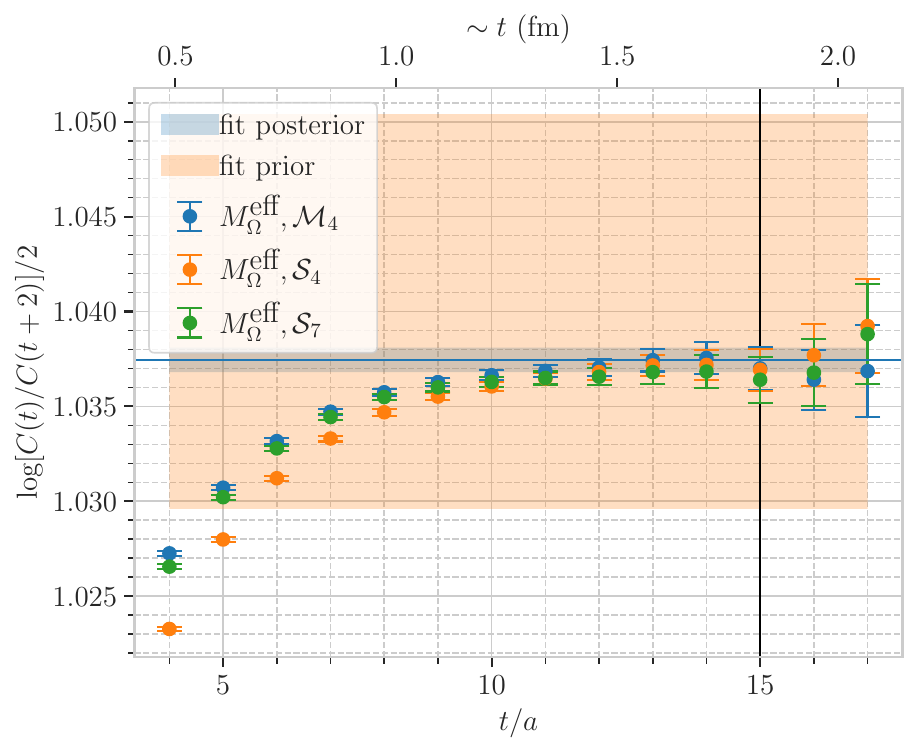}
    \caption{The three tastes of the $\Omega$ baryon (the two \omegaones\ multiplets, interpolated by $\mathcal{S}_4$ and $\mathcal{S}_7$, and the single \omegatwos\ multiplet, interpolated by $\mathcal{M}_4$) on the $a=0.12$~fm ensemble.
    Effective masses have been averaged in time via the method in Ref.~\cite{Bailey:2008wp} to suppress oscillatory states.
    The blue horizontal band shows the Bayesian model averaged fit result to the \omegatwos\ taste.
    The prior width (yellow) is an order of magnitude larger than the posterior fit width, indicating that the prior is nonconstraining.
    The vertical line indicates the largest value of $t$ used in the fit.}
    \label{fig:eff-mass-prior}
\end{figure}
Figure~\ref{fig:eff-mass-prior} shows an example fit with the much larger prior for comparison.

The maximum time $t_\text{max}$ in the fit ranges used was fixed at about 1.8~fm, corresponding to a statistical uncertainty in $C(t_\text{max})$ of about 0.4\% on most ensembles.
We then performed a Bayesian model average (BMA)~\cite{Jay:2020jkz, Neil:2022joj} over both the number of states in the fit (comparing $1+1$- and $2+2$-state fits) and the minimum time $t_\text{min}$ (varied from about 0.5~fm to $t_\text{max}-5a$, so all fits have at least six data points) using the procedure discussed in Appendix~\ref{app:BMA}.
All fit parameters are treated as free parameters that count against the total number of degrees of freedom since the prior widths are large enough to be unconstraining.
The results of this procedure are tabulated in Appendix~\ref{app:lattice-results} and used in subsequent fits.
As a cross-check, fits with additional states ($3+3$- and $4+4$-state fits) have also been performed; these are consistent with the $1+1$- and $2+2$-state fits on coarser ensembles but numerically unstable on finer ensembles (with fewer configurations) and are therefore excluded from the final averages.

\begin{figure}
    \centering
    \includegraphics[width=\linewidth]{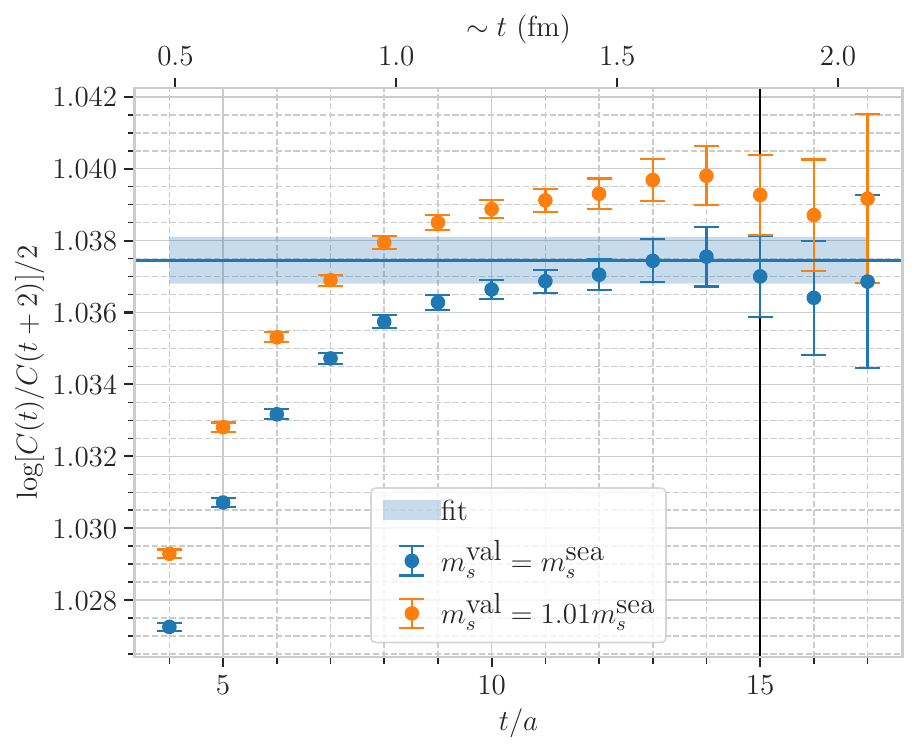}

    \caption{The variation in the $\Omega$ baryon mass with $m_s^\text{val}$.  Shown are the unitary point $m_s^\text{val} = m_s^\text{sea}$ and a slightly heavier valence mass $m_s^\text{val}= 1.01 m_s^\text{sea}$.  While the uncertainty on a single-mass fit is comparable to the difference between the two $\Omega$ baryon masses, the two correlation functions are highly correlated (and effective masses fluctuate in tandem), so the correlated difference between the $\Omega$ baryon masses can be resolved to few-percent accuracy.}
    \label{fig:eff-mass-valence}
\end{figure}
On ensembles with multiple valence strange-quark masses used to study the quark-mass dependence of $M_\Omega$ (as shown in Fig.~\ref{fig:eff-mass-valence}), we perform a joint fit to the pair of correlation functions.
The results of the joint fits are combined using a Bayesian model average with the same prescription as single fits (albeit with twice as many degrees of freedom from the two correlation functions).
This procedure furnishes a model-averaged correlated difference between the masses.
Strong correlations between the input hadronic correlation functions result in precise values for the correlated differences and therefore also the slope of $M_\Omega$ versus $m_s$.
As the individual correlation functions are highly correlated, this correlated difference (and therefore the slope of $M_\Omega$ versus $m_s$) is very precise.

As a cross-check, we have also computed positive-definite correlation functions with identically smeared source and sink.
The masses obtained are consistent with those from point sources but with statistical uncertainties too large to be meaningfully constraining.

Lanczos/Rayleigh-Ritz methods~\cite{Wagman:2024rid,Hackett:2024xnx,Ostmeyer:2024qgu,Chakraborty:2024scw,Hackett:2024nbe,Abbott:2025yhm} that have recently been developed provide further cross-checks of our correlator fits.
When this approach is applied to $C(t)$ with $t \in [t_{\rm min}, t_{\rm min} + 2m-1]$, algebraic energy estimators $E_i = -\ln\lambda_i$ are first obtained from Ritz values $\lambda_i$, which can be computed by solving generalized eigenvalue problems $(H^{(0)}, H^{(1)})$ of Hankel matrices $H^{(p)}_{ij} \equiv C(t_{\rm min} + i + j + p)$ as in the generalized-pencil-of-functions method~\cite{Hua:1989,Sarkar:1995,Aubin:2010jc,Aubin:2011zz}.
The Ritz values and associated overlap factors $c_i$ provide an exact $m$-state decomposition of the correlator $f(t) = \sum_{i=1}^m c_i \lambda_i^t$ for this time range.
Subsequently, spurious-state filtering is applied to remove spurious Ritz values arising from statistical noise using either the Cullum-Willoughby (CW) test~\cite{Cullum:1981,Cullum:1985} or the $Z$-factor Cullum-Willoughby (ZCW) test~\cite{Hackett:2024nbe}.

The first step, computing Ritz values, does not require any modifications for staggered fermions~\cite{Wagman:2024rid}.
However, ZCW filtering does require slight modification due to negative Ritz values and squared overlaps arising from oscillating states.
In particular, the cutoff $\varepsilon_{\rm ZCW}$ for the minimum overlap of a nonspurious state can be set based on the last iteration where all Ritz values and overlaps are in their physical ranges~\cite{Hackett:2024nbe}.  For staggered fermions, one should instead consider the last iteration where at least half the states satisfy all physical constraints,  as the other half are  oscillating states corresponding to negative eigenvalues of the transfer matrix.
We then set $\varepsilon_{\rm ZCW}$ to the minimum normalized overlap product $|c_i/C(t_{\rm min})|$ from this iteration divided by a hyperparameter $F_{\rm ZCW}$.
After filtering, the ground-state energy estimator is defined as minus the logarithm of the largest Ritz value.
To avoid outliers arising from spurious state misidentification, we use bootstrap median estimators whose uncertainties are computed with a nested bootstrap approach~\cite{Wagman:2024rid,Hackett:2024xnx,Ostmeyer:2024qgu,Chakraborty:2024scw,Hackett:2024nbe}.

\begin{figure}
    \centering
    \includegraphics[width=\linewidth]{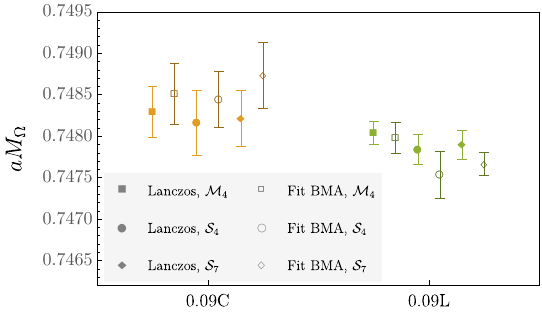}
    \caption{Comparison of filtered Lanczos/Rayleigh-Ritz and BMA correlator fit results for the two retuned $a=0.09$ fm ensembles, which differ only in their physical volume, and all three $\Omega$ baryon tastes.}
    \label{fig:lanczos-comp}
\end{figure}

We apply this method to all correlator results with $t \in [2, N_t / 2]$ and quote the median energy estimator from the last three iterations as a final result (large-iteration results are highly correlated).
Filtered Lanczos/Rayleigh-Ritz results using the ZCW test with $F_{\rm ZCW} = 10$ are statistically consistent with the correlator fits described above.
Similar uncertainties are obtained by filtered Lanczos/Rayleigh-Ritz and correlator fits.
Consistent results with similar uncertainties are also obtained using the CW test with the cutoff $\varepsilon_{\rm CW}$ chosen as described in Ref.~\cite{Wagman:2024rid} with default hyperparameters described therein.
While these Lanczos fits were not included in subsequent-stage fits, they are in good agreement with BMA correlator results, as shown in Fig.~\ref{fig:lanczos-comp} for the $a\approx 0.09$ fm ensembles.

\subsection{Electromagnetic corrections}
\label{sec:pure-QCD-MOmega}

Let $M_\Omega(m_v,q_v)$ be the mass of the spin-$3/2$\ baryon made from three identical quarks of bare valence mass $m_v$ and electric charge~$eq_v$, where $e$ is the charge of the positron.
We are interested in the electromagnetic contribution to $M_\Omega(m_v,q_v)$,
\begin{equation}
    \emma{M_\Omega} \equiv M_\Omega(\zeta m_s, -\third) - M_\Omega(m_s, 0).
    \label{eq:emma}
\end{equation}
When the electric charge is nonzero, the additional ultraviolet divergences of QED require a different tuning of the valence mass, 
and $\zeta m_s$ is the bare mass corresponding to the strange quark in QCD+QED, while $m_s$ denotes the bare mass corresponding to the strange quark in pure QCD.
The precise definition of $\zeta$ is connected to the scheme for matching pure (electrically neutral) QCD to the real world of QCD+QED.

In chiral perturbation theory ($\chi$PT) combined with QED, the pseudoscalar meson masses satisfy
\begin{equation}
    M^2_{xy} = B_0 (m_x+m_y) + e^2 \Delta_\text{EM} (q_x-q_y)^2 + \cdots ,
\end{equation}
where $x$ and $y$ label the flavors of the valence quark and antiquark and the $\cdots$ denote higher orders in $\chi$PT+QED.
This motivates the definition
\begin{align}
    v(m_v, q_v) &\equiv M^2_{K^0} + M^2_{K^+} - M^2_{\pi^+} \nonumber \\
        &= 2B_0 m_v + 2e^2\Delta_\text{EM} (q_v-q_u)(q_v-q_d) + \cdots \nonumber \\
        &= 2 B_0 m_v + \cdots
    \label{eq:v-definition}
\end{align}
at valence strange quark mass $m_v$, which does not depend on $m_u$ or $m_d$ until next-to-leading order (NLO).
Since $q_s=q_d$, $v$ is independent of the charges until NLO as well.

The factor $\zeta$ can now be fixed by adjusting the quantities $v(\zeta m_s,-\third)$ and $v(m_s,0)$, respectively, to the experimental QCD+QED value in the Particle Data Group \cite{ParticleDataGroup:2024cfk} and the pure-QCD FLAG average value \cite{FlavourLatticeAveragingGroupFLAG:2024oxs}
\begin{align}
    v(\zeta m_s,-\third) = v_\text{PDG}  &= 4.7186(2) \times 10^5~\text{MeV}^2 , \\
    v(m_s, 0)            = v_\text{FLAG} &= 4.7103332 \times 10^5~\text{MeV}^2 ,
\end{align}
where the difference between these quantities is indicative of NLO electromagnetic effects in Eq.~(\ref{eq:v-definition}).
The dependence on the valence strange quark mass can be approximated as
\begin{equation}
    v(\zeta m_s, q_v) \approx \zeta v(m_s, q_v) \, ,
\end{equation}
so
\begin{align}
   \zeta &= \frac{v(\zeta m_s, -\third)}{v(m_s, -\third)} \nonumber \\
   &= \frac{v(m_s, 0)}{v(m_s, -\third)} \frac{v(\zeta m_s, -\third)}{v(m_s, 0)} \nonumber \\
   &= \frac{a^2 v(m_s, 0)}{a^2 v(m_s, -\third)}\frac{v_\text{PDG}}{v_\text{FLAG}} ,
   \label{eq:zeta}
\end{align}
where the lattice spacing has been inserted into the first factor as a reminder that it is a ratio of lattice calculations, while 
the second factor is the ratio of strange-quark-mass tuning conditions.
Since the factors of $a^2$ cancel (as $w_0/a$ and $w_0$ are independent of valence quark charges), $\zeta$ can be obtained without any prescription for the conversion from lattice units to physical units.

It is convenient to abbreviate the mass shift at fixed bare quark mass
\begin{equation}
    \Delta M_\Omega(m_v) \equiv M_\Omega(m_v, -\third) - M_\Omega(m_v, 0) ,
    \label{eq:em-bare}
\end{equation}
and we can take the correlated difference between data at the sea strange quark mass $m_v=m'_s\equiv m^\text{sea}_s$.
This was computed using strange-quark propagators on $\text{SU}(3)\times\text{U}(1)$ configurations, as discussed in Sec.~\ref{sec:em-setup}, with U(1) charge $q=\pm\third$.
After averaging over the positive and negative charges and over the source locations, we obtain the difference in Eq.~(\ref{eq:em-bare}) above.

Quark-mass mistuning is accommodated by expanding Eq.~(\ref{eq:emma}) around $m'_s$:
\begin{align}
    \emma{M_\Omega} &= \Delta M_\Omega(m'_s) + (\zeta m_s -m'_s) \frac{M_\Omega}{m'_s} \frac{\partial\ln M_\Omega}{\partial\ln m'_s} \nonumber \\
           &\qquad - (m_s -m'_s) \frac{M_\Omega}{m'_s} \frac{\partial\ln M_\Omega}{\partial\ln m'_s} \nonumber \\
       &=  \Delta M_\Omega(m'_s) + (\zeta-1) M_\Omega \frac{\partial\ln M_\Omega}{\partial\ln m'_s}
    \label{eq:emma-e}
\end{align}
neglecting higher-order terms.\footnote{The distinction between $M_\Omega(q=0)$ and $M_\Omega(q=-1/3)$ in the term multiplying $(\zeta-1)$ is higher order in $\alpha$ and the strange mass mistuning $m_s-m_s'$ and is negligible at the precision of this calculation.}
Now, $\zeta-1 \sim q_s^2\alpha\ln am'_s$ grows as $a\to0$, so eventually this approximation fails, but even on the finest lattice, where $a \approx 0.04$~fm, $-\ln am'_s\approx4$ is not yet too big.
For taking the continuum limit of $\emma{M_\Omega}$, it is natural to work with the relative contribution,
\begin{equation}
    \frac{\emma{M_\Omega}}{M_\Omega} = \frac{\Delta M_\Omega(m'_s)}{M_\Omega} + (\zeta-1) \frac{\partial\ln M_\Omega}{\partial\ln m'_s}
\end{equation}
with $\zeta$ from Eq.~(\ref{eq:zeta}) and
\begin{equation}
    \frac{\partial \ln M_\Omega}{\partial \ln m_s'} \approx \frac{\partial \ln M_\Omega}{\partial \ln v} = 0.24(1)    
\end{equation}
taken from pure-QCD global fits below as discussed in Appendix~\ref{app:stuff}.

When computing $M_\Omega(m_v,q_v)$ for nonzero $q_v$, there are power-law (not exponentially) suppressed finite-volume effects that have to be corrected for in both the $\Omega$ baryon and meson masses before combining data for Eq.~(\ref{eq:emma}) into a continuum limit.
Finite-volume corrections $\delta_\text{EM}^\text{FV} M$ to the mass $M$ of any charged hadron can be estimated using the point-particle approximation \cite{BMW:2014pzb} truncated at N$^2$LO, before structure-dependent corrections enter, namely
\begin{equation}
    \frac{\delta_\text{EM}^\text{FV} M}{M} = \frac{ \kappa q^2 \alpha_\text{EM}}{2 M L} \left[ 1 + \frac{2}{M L} \left( 1 - \frac{\pi}{2\kappa}\frac{T}{L} \right) \right]
    \label{em-fv}
\end{equation}
with $\kappa \approx 2.837297$ defined via an integral expression in Ref.~\cite{BMW:2014pzb} and $q$ the hadron charge. 
Note that unlike the finite-volume corrections to the pure-QCD mass, which are exponentially suppressed and subdominant, corrections to $\Delta M$ are only power-law suppressed and cannot be altogether neglected.
The small difference between NLO and N$^2$LO finite-volume corrections is taken to be a systematic uncertainty on $\delta_\text{EM}^\text{FV} M$.

The relative correction $\emma{M_\Omega}/M_\Omega$ was computed on the three coarsest ensembles in this work (0.15, 0.12, and 0.09C) and then extrapolated to the continuum using a fit function linear in $a^2 \alpha_s$, the leading-order contribution to discretization effects for hadron masses measured using the HISQ action.
\begin{figure}
    \centering
    \includegraphics[width=\linewidth]{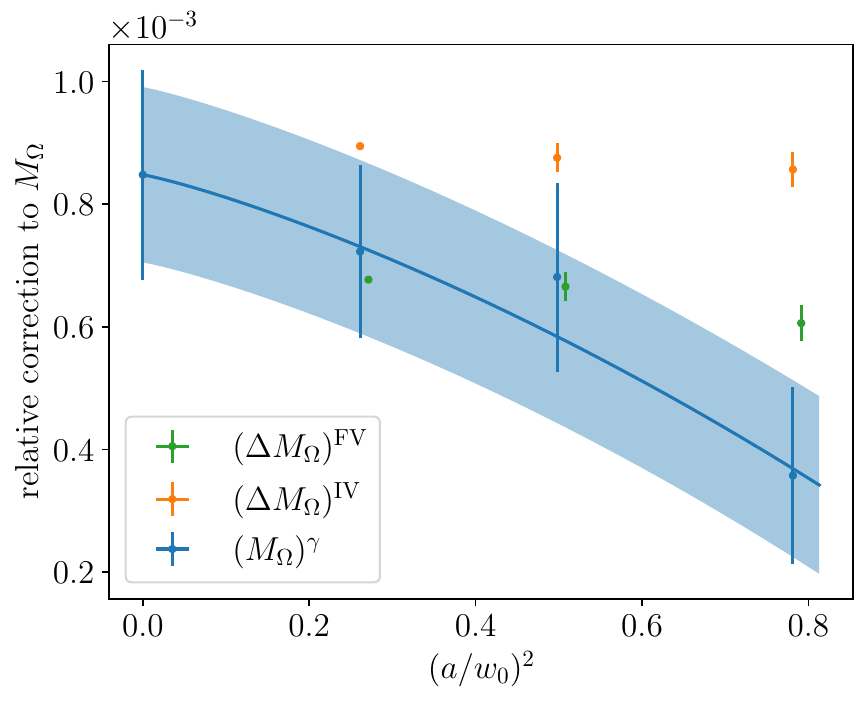}
    \caption{
    Relative size of the electromagnetic contribution to the $\Omega^-$ mass.  The green points represent the difference in the $\Omega^-$ mass in QCD+QED versus pure QCD at fixed bare quark masses $\Delta M_\Omega(m'_s)$ with no corrections applied (but a small horizontal offset for clarity).  The orange points include finite-volume corrections to electromagnetic effects from Eq.~(\ref{em-fv}).  The blue points additionally include the effects of quark mass retuning from Eq.~(\ref{eq:zeta}), including the uncertainty from electroquenching in the meson sector.  The blue curve is the continuum extrapolation of $\emma{M_\Omega} / M_\Omega$ with all corrections applied, and the continuum limit includes an additional uncertainty from electroquenching in the baryon sector.
    The case of the \omegatwos\ state with the SSCc definition of $w_0/a$ is shown; the flavor-symmetric \omegaones\ states and other definitions of $w_0/a$ give results consistent within a small fraction of the electroquenching error.}
    \label{fig:em-extrapolation}
\end{figure}
The extrapolation for the mixed-symmetry, strange-isospin \omegatwos\ baryon is shown in Fig.~\ref{fig:em-extrapolation},
giving a continuum-limit value of
\begin{equation}
    \frac{\emma{M_\Omega}}{M_\Omega} = 8.5(1.7) \times 10^{-4}.
    \label{eq:emmaOmega}
\end{equation}
The result in Eq.~(\ref{eq:emmaOmega}) has been obtained in the electroquenched approximation, in which sea quark charges are neglected.
Errors from this approximation enter both indirectly through the meson masses (and in turn the renormalization factor $\zeta$) as well as directly in sea-quark diagrams contributing to $M_\Omega$.
The size of the missing contributions can be estimated in the $1/N_c$ approximation~\cite{Chow:1995by}.

For mesons, the ratio $v(\zeta m_s,-\third)/v(m_s,0)$ differs from its quenched counterpart by effects that one expects to be approximately the valence contribution (i.e., the difference from 1) divided by $N_c$.  Thus, one can approximate the uncertainty here by $\left[v(\zeta m_s,-\third)/v(m_s,0) - 1 \right]/N_c$, which we add as a systematic error on $\zeta$.

For baryons, the leading-order QED effects of sea quarks enter through sea-sea diagrams (with the photon attached to two sea quarks) and sea-valence diagrams (with one photon end attached to a sea quark and the other attached to a valence quark).  Since the baryon-to-meson electric charge ratio grows like $N_c$, sea-sea electromagnetic corrections to baryon masses are suppressed by two powers of $N_c$.  Sea-valence diagrams are suppressed by only one power of $N_c$ but are zero in the $\text{SU}(3)$ flavor-symmetric limit (since $q_u + q_d + q_s = 0$), so they are also suppressed by meson mass differences that scale as $(M_K^2 - M_\pi^2)/(4\pi f_\pi)^2 \sim 0.2$.  One can then take $\max(1/N_c^2, 0.2/N_c) = 1/N_c^2 = 1/9$ as an additional systematic uncertainty in $(M_\Omega)^\gamma$.

Electroquenching is the dominant systematic uncertainty in the value of $(M_\Omega)^\gamma$ in Eq.~(\ref{eq:emmaOmega}), and electrodynamical calculations in subsequent work would be beneficial for removing this systematic.
However, since electromagnetism is a small effect in absolute terms, this is a small fraction of the total error budget in the final determination of $w_0$ below.

Subtracting $(M_\Omega)^\gamma$ from the PDG average value of the experimental (i.e., QCD+QED) $\Omega^-$
mass of 1672.43(32)~MeV~\cite{ParticleDataGroup:2024cfk}%
\footnote{The PDG lists an average of 1672.43(32)~MeV for all measurements of the $\Omega^-$ baryon and an average of 1672.5(7)~MeV for measurements of its antiparticle $\bar\Omega^+$, giving a global fit of 1672.45(29)~MeV assuming CPT symmetry.
This work uses the former value, averaging only $\Omega^-$ data, as in Ref.~\cite{Miller:2020evg}.
The difference is negligible at the current precision, corresponding to a change of about 0.02$\sigma$ in the extracted $w_0$ value.}
gives a pure-QCD $\Omega$ mass of
\begin{equation}
    M_\Omega = 1671.01(43)~\text{MeV} \, . 
    \label{eq:momega-qcd}
\end{equation}
Since the $\Omega$ baryon has isospin~0, no correction for isospin-breaking is necessary until second order, which is beyond the precision needed for Eq.~(\ref{eq:momega-qcd}).
Therefore, this can be used on equal footing as the isospin-symmetric pure-QCD meson masses in Ref.~\cite{FlavourLatticeAveragingGroupFLAG:2024oxs}
\begin{align}
    M_\pi &= 135.0~\text{MeV} , \label{mpi-qcd} \\
    M_K   &= 494.6~\text{MeV} , \label{mK-qcd}
\end{align}
to define a pure-QCD world that serves as inputs for the determination of $w_0$ in pure QCD.
This scheme is equivalent to the FLAG scheme~\cite{FlavourLatticeAveragingGroupFLAG:2024oxs} (sometimes called the Edinburgh consensus) except that $M_\Omega$ is used for scale setting instead of the pion decay constant~$f_\pi$.

\subsection{Chiral-continuum fits}
\label{sec:w0MOmega}

In a pure-QCD simulation, extrapolation to the physical point requires meson masses to be matched to their electromagnetism-subtracted values given in Eqs.~(\ref{eq:momega-qcd})--(\ref{mK-qcd}).
With the global scale not known \emph{a priori} in this calculation, one must solve self-consistently for both meson masses and the overall scale from computed values of $aM_K$, $aM_\pi$, $aM_\Omega$ in lattice units.
Following Ref.~\cite{Miller:2020evg}, we form the dimensionless ratios $M_\pi/M_\Omega$ and $M_K/M_\Omega$ and adjust the quark masses to obtain the ratios implied by Eqs.~(\ref{eq:momega-qcd})--(\ref{mK-qcd}).
To avoid analyst bias, the pure-QCD $\Omega$ baryon mass in Eq.~(\ref{eq:momega-qcd}) was multiplied by a blinding factor chosen from a Gaussian with width 0.01 until the analysis was frozen.

To leading order in $\chi$PT, the linear combinations of meson masses proportional to the strange valence mass and the sum of the sea quark masses are\footnote{Note that this work uses $s_\Omega$ for the sum of the sea masses rather than the strange mass as in Ref.~\cite{Miller:2020evg} and also that this work does not square $v_\Omega$ and $s_\Omega$ in the definitions.}
\begin{align}
    v_\Omega &\equiv \frac{2(M_K^\text{val})^2 - (M_\pi^\text{val})^2}{M_\Omega^2}  , \\
    s_\Omega &\equiv \frac{2(M_K^\text{sea})^2 + (M_\pi^\text{sea})^2}{M_\Omega^2}  .
\end{align}
Inserting Eqs.~(\ref{eq:momega-qcd})--(\ref{mK-qcd}) yields the physical values
\begin{align}
    v_\Omega^\text{phys} &= 0.16869(9) , \\
    s_\Omega^\text{phys} &= 0.18175(9)  ,
\end{align}
where $v_\Omega$ is the linear combination of meson masses $v$ in Eq.~(\ref{eq:v-definition}) divided by $M_\Omega^2$.

\begin{widetext}
About the quark-mass corrected point $(w_0 M_\Omega)^\text{corr}$, one can expand $w_0 M_\Omega$ through linear order in the deviations of quark masses from their physical values as
\begin{equation}
    (w_0 M_\Omega)^\text{latt} = (w_0 M_\Omega)^\text{corr} \left[ 1 + \gamma_v \frac{v_\Omega^\text{latt} - v_\Omega^\text{phys}}{v_\Omega^\text{phys}} + \gamma_s \frac{s_\Omega^\text{latt} - s_\Omega^\text{phys}}{s_\Omega^\text{phys}} + \cdots \right] ,
\end{equation}
for some quark-mass independent coefficients $\gamma_v$, $\gamma_s$.
Both $\gamma_v$ and $\gamma_s$ are nonzero at leading order in $\chi$PT, although the results of this fit find $\gamma_v$ numerically larger than $\gamma_s$ by a factor of about~5.

The mass-corrected values of $w_0 M_\Omega$ are related to their continuum values via the expansion
\begin{equation}
    (w_0 M_\Omega) (a) = (w_0 M_\Omega)^\text{cont}\left[1 + c_1 (a/w_0)^2 \alpha_s (a)^{\{0,1\}} + c_2 (a/w_0)^4 + c_3 (a/w_0)^6 + \cdots\right] ,
    \label{eq:continuum-extrapolation}
\end{equation}
defining the lattice spacing explicitly via (one of the definitions of) $w_0/a$.
Only even powers of $a$ appear because staggered fermions obey a taste nonsinglet chiral symmetry.

The improvement to staggered fermions from the Naik term suppresses the $a^2$ term in $aM_\Omega$ by a factor of $\alpha_s$, which approaches zero logarithmically in the continuum limit.
Discretization effects in the flow scale $w_0/a$, however, begin at $O(a^2)$ with no additional factors of $\alpha_s(a)$.
As a result, fits with and without this power of $\alpha_s$ are included in the BMA.%
\footnote{At the time of unblinding, we did not appreciate the fact that $w_0/a$ does not include a factor of $\alpha_s (a)$, even in schemes where the flow is improved.  Until then, only fits including the factor of $\alpha_s$ were included in the BMA.
In post-unblinding discussions, this oversight was pointed out \cite{private-comm}, after which we began to incorporate fits without $\alpha_s$ into the BMA.
This change shifted the original result for $w_0$ by about 0.1$\sigma$ and increased the uncertainty by about 10\%.}

At finite lattice spacing, the quark-mass corrections can also depend on the lattice spacing, so one can write, for example,
\begin{equation}
    \gamma_v (a) = \gamma_v^{(0)} \left[ 1 + \gamma_v^{(1)} (a/w_0)^2 \alpha_s^{\{0,1\}} + \gamma_v^{(2)} (a/w_0)^4 + \cdots \right]
\end{equation}
and similarly for $\gamma_s$.
However, since $\gamma_s$ is numerically smaller than $\gamma_v$, and the computed pion-mass dependence of $M_\Omega$ at $a=0.12$~fm and $a=0.09$~fm are consistent (within $\sim 50\%$ uncertainties), discretization effects in $\gamma_s$ are neglected,\footnote{As a cross-check, we also performed the BMA including fits with the term $\gamma_s^{(1)} a^2 \alpha_s$; this shifted the central value by 0.02$\sigma$ and left the uncertainty essentially unchanged.} and this work sets $\gamma_s(a)=\gamma_s^{(0)}$.
Assembling all the pieces, one has
\begin{equation}
    \begin{split}
        (w_0 M_\Omega)^\text{latt} (a, m_\text{val}, m_\text{sea}) &= (w_0 M_\Omega)^\text{cont}\bigg\{ 1 + c_1 (a/w_0)^2 \alpha_s (a)^{\{0,1\}} + c_2 (a/w_0)^4 + c_3 (a/w_0)^6  \\
        &+ \gamma_v^{(0)} \left[ 1 + \gamma_v^{(1)} (a/w_0)^2 \alpha_s (a)^{\{0,1\}} 
        + \gamma_v^{(2)} (a/w_0)^4 \right] \frac{v_\Omega^\text{latt} - v_\Omega^\text{phys}}{v_\Omega^\text{phys}} +
        \gamma_s^{(0)} \frac{s_\Omega^\text{latt} - s_\Omega^\text{phys}}{s_\Omega^\text{phys}}\bigg\} .
    \end{split}
    \label{eq:continuum-extrapolation-full}
\end{equation}
\end{widetext}
The lattice spacing $a$ appearing as the argument of~$\alpha_s$ must also be made explicit, and we choose
\begin{equation}
    a  \equiv \frac{(w_0 M_\Omega)^\text{cont}}{M_\Omega^\text{phys}}\frac{a}{w_0}
\end{equation}
so that the right side of Eq.~(\ref{eq:continuum-extrapolation-full}) is a (nonlinear) function of the fit coefficients $(w_0 M_\Omega)^\text{cont}$, $c_1$, $c_2$, $c_3$, \ldots, the calculated values of $w_0/a$ on the various ensembles, and the quark mass corrections.
The value of $\alpha_s$ as a function of lattice spacing was obtained by running from $\alpha_s (5~\text{GeV}) = 0.2530$ \cite{Chakraborty:2014aca} to the scale of $2/a$ at four-loop order using the package \verb|qcdevol|~\cite{qcdevol}.

As with the electromagnetic correction, we apply the BMA procedure to obtain a final result for $(w_0 M_\Omega)^\text{cont}$.
To assess the impact of discretization effects, we perform the following variations in the fit function and data set:
\begin{itemize}
    \item order of the discretization effects to the continuum extrapolation: $\{c_1\}$, $\{c_1, c_2\}$, $\{c_1, c_2, c_3\}$;
    \item order of the discretization effects to the valence-mass correction: $\{\gamma_v^{(0)}\}$, $\{\gamma_v^{(0)}, \gamma_v^{(1)}\}$, $\{\gamma_v^{(0)}, \gamma_v^{(1)}, \gamma_v^{(2)}\}$;
    \item range of data included: $[0.042, 0.15]$~fm, $[0.042, 0.12]$~fm, $[0.042, 0.09]$~fm.
\end{itemize}
Because the maximum sizes of the $\{c_i\}$ and $\{\gamma_v^{(i)}\}$ sets are reduced when excluding coarser lattices, the variations above correspond to a total of 14 physical-point fit variations.
As above, we also repeat the physical-point extrapolation for the twelve definitions of $w_0/a$ and three tastes of interpolating operators used for $a M_\Omega$.
In total, this gives 36 possible datasets times 14 fit functions that are averaged to the final continuum-limit result for~$w_0M_\Omega$.

\begin{figure*}
    \centering
    \includegraphics[width=0.49\linewidth]{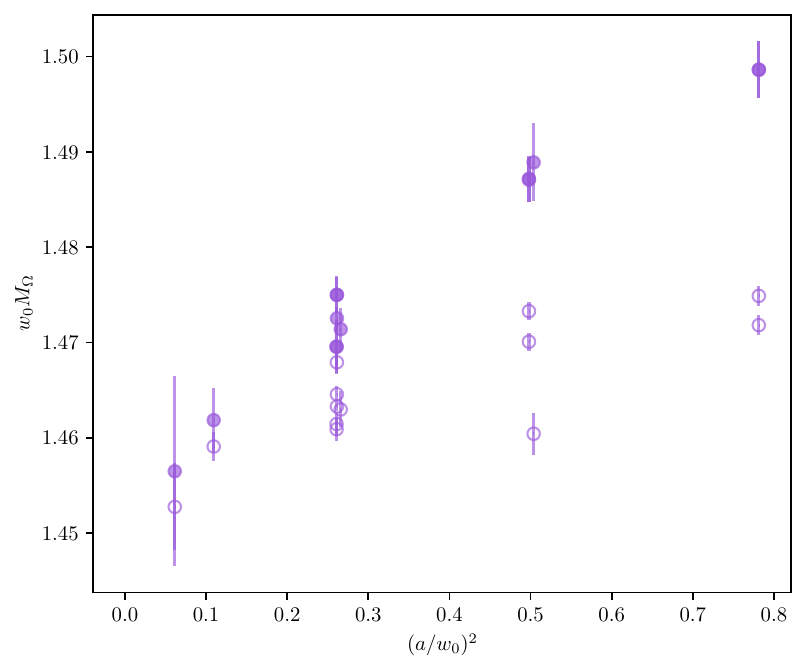}
    \includegraphics[width=0.49\linewidth]{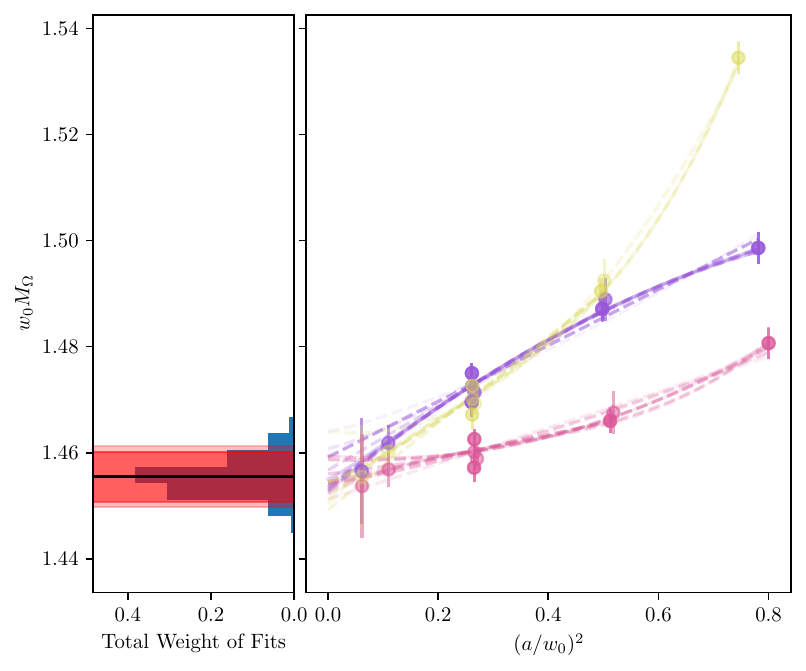}
    \caption{
    (Left) The impact of corrections for quark mass retuning on $M_\Omega$.
    The uncorrected values of $aM_\Omega$ (using the \omegatwos\ interpolating operators), multiplied by the corresponding $w_0/a$ values (in one of the twelve schemes for concreteness), are shown in open circles on the nine ensembles.
    The filled circles show quark-mass-corrected data points, with the corrections applied using the highest-weight fit.  Note that at a given lattice spacing, applying the quark mass corrections significantly improves the agreement among various sea and valence masses but also inflates the uncertainties, as discussed in Appendix~\ref{app:stuff}.
    (Right) The various extrapolations of $w_0 M_\Omega$ to the continuum.
    The left panel shows a histogram in blue of the continuum-limit values of $w_0 M_\Omega$, weighted by the expression in Eq.~(\ref{eq:model-weights}) applied to the fit generating that value.
    Overlaid on the histogram are a black line representing the central value of $(w_0 M_\Omega)^\text{cont}$ and red and pink bands showing its statistical and total uncertainties.
    The right panel shows representative fit curves from three $w_0/a$ definitions (colored to match the $w_0/a$ definitions listed in Fig.~\ref{fig:continuum-extrapolation-summary}) and the 14 physical-point extrapolation models shaded proportionally to their model weights, with the model with the highest overall weight (quadratic fit in $a^2$ with $v_\Omega a^2$ mass corrections to all lattice spacings) for each $w_0/a$ definition shown with a solid line.}
    \label{fig:continuum-extrapolation-multiple}
\end{figure*}
\begin{figure*}
    \centering
    \includegraphics[height=0.78\textheight]{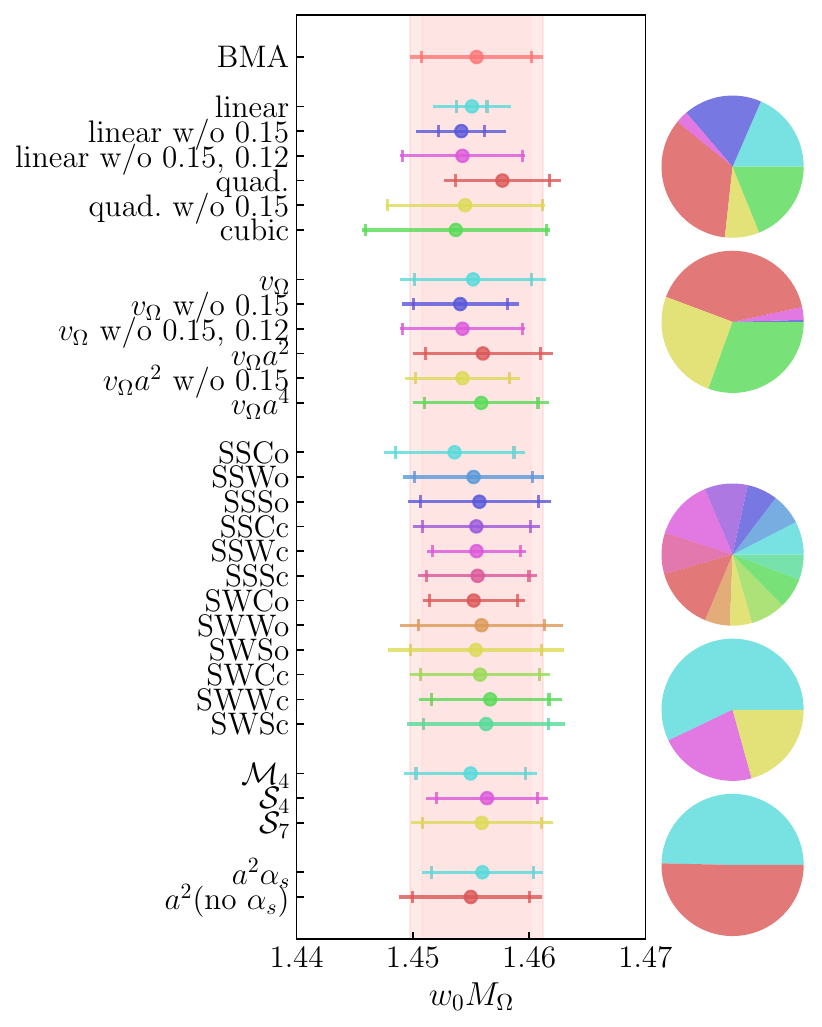}
    \caption{Summary of model variations in the fits used to extract $w_0 M_\Omega$ at the physical point.  The different groupings represent the order of the continuum extrapolation, the order of discretization effects included in the valence-mass correction based on $v_\Omega$, the gradient flow used (Symanzik or Wilson for the flow action, improved or unimproved, and clover, Wilson, or Symanzik for the flow observable), the representation of the $\Omega$ baryon used (mixed representation or subgroups 4 or 7 of the symmetric representation), and the power of $\alpha_s$ in the leading term in the continuum extrapolation.  Inner error bars represent statistical uncertainties and outer bars also include systematic uncertainties from varying other submodels.  The pie charts on the right represent the relative weight of each submodel within the corresponding class of variations.}
    \label{fig:continuum-extrapolation-summary}
\end{figure*}

Some representative fits are shown in Fig.~\ref{fig:continuum-extrapolation-multiple}, with the full set of these variations, as well as the changes in fit model, shown in Fig.~\ref{fig:continuum-extrapolation-summary}.
For each class of submodels (namely, variations on the details of the continuum extrapolation, the mass corrections, the $w_0/a$ discretization, the $\Omega$ baryon taste, and the power of $\alpha_s$), a set of results is shown for the various choices within that submodel with Bayesian model averaging over the remaining submodels.
The full space of choices within the different submodels weighted by the corresponding subset probabilities is used in the global BMA~\cite{FermilabLatticeHPQCD:2023jof}.
Variation among different models and datasets is relatively mild, and the uncertainty budget is dominated by the statistical uncertainty in the data.

\subsection{Uncertainty budget}

In this section, we present estimates of the contributions of various statistical and systematic effects to the final quoted uncertainty of $0.4\%$ in $w_0$ in Eq.~(\ref{eq:w0-final}).  These are summarized in Table~\ref{tab:error-budget}.  Global fit systematics reflect the deviations of the various submodels from the model average, as prescribed in Ref.~\cite{Jay:2020jkz}.  The other uncertainties can be approximated by turning off the corresponding uncertainty in the input to the final-stage fit, keeping all submodel weights fixed, and determining the change in variance that would arise from eliminating that source of uncertainty.  As a cross-check, the contributions from the various components sum in quadrature to about 0.39\%, in good agreement with the total 0.40\% uncertainty.

The dominant uncertainties are statisical errors in the fits to correlation functions and systematic errors in the extrapolation to the physical point.  These can be further reduced by additional data, especially on the finest $a=0.04$~fm ensemble.

QCD finite-volume corrections on $M_\Omega$ are negligible: the contribution from kaon loops is proportional to $\exp(-M_KL)\lesssim\exp(-12)<10^{-5}$, and
contributions from pion loops are proportional to
$(M_\pi/4\pi f_\pi)^4 e^{-M_\pi L}/(M_\pi L)^{3/2} \approx 10^{-6}$~\cite{Miller:2020evg}.
Statistical consistency between the QCD contributions on the two volumes used at 0.09~fm confirms this prediction of negligible finite-volume corrections, albeit at the level of $10^{-3}$, which can be seen comparing the two volumes shown in Fig.~\ref{fig:lanczos-comp}.
Finite-volume corrections to electromagnetism are larger but are well controlled by Eq.~(\ref{em-fv}); truncation errors from this approximation are included in the electromagnetism uncertainty but are much smaller than the electroquenching error.
Strong-isospin breaking effects are also negligible, as the $\Omega$ baryon is isospin-0, so effects only enter at second order in $(m_u - m_d)/\Lambda_\text{QCD}$ and are therefore estimated to be $\approx 10^{-4}$.

\begin{table}
    \centering
    \caption{Estimates of the approximate impact of various uncertainties on the total uncertainty budget.
    Quantities are shown as a percentage of the final quoted value of $w_0$ (fm).}
    \label{tab:error-budget}
    \begin{tabular}{cc}
        \hline\hline
            Source                           & Uncertainty (\%) \\ 
        \hline
            Correlator fits (stat.~$+$ sys.) & 0.32 \\
            Gradient flow (stat.)            & 0.04 \\
            Experimental hadron masses       & 0.02 \\
            Electroquenching                 & 0.02 \\
            Global fit systematics           & 0.22 \\
        \hline
            Total                            & 0.40 \\ 
        \hline\hline
    \end{tabular}
\end{table}

\section{Conclusions and Outlook}
\label{sec:conclusions}

We obtain from the analysis of the previous section the following QCD value for the $\Omega$ baryon mass in units of the gradient-flow scale:
\begin{align}
    w_0 M_\Omega &= 1.4555(47)(32)[57] .
    \label{eq:w0momega-final}
\end{align}
The error in the first parentheses is from correlator fits (statistics and fit systematics), while the error in the second parentheses is from all other systematics.
The final error in brackets is their sum in quadrature and constitutes the total error from the BMA over chiral-continuum fit variations.
Combining Eq.~(\ref{eq:w0momega-final}) with our QCD value of $M_\Omega=1671.01(43)$~MeV [Eq.~(\ref{eq:momega-qcd})], we obtain for the physical value of the gradient-flow scale
\begin{align}
    w_0 &= 0.17187(56)(38)[68]~\text{fm} ,
    \label{eq:w0-final}
\end{align}
with the errors in parentheses and brackets defined analogously as before.
This is our result in both QCD+QED and pure QCD.
The electroquenched approximation used here is insensitive to the difference, and the uncertainty covers the error from electroquenching.

\begin{figure}
    \centering
    \includegraphics[width=\linewidth]{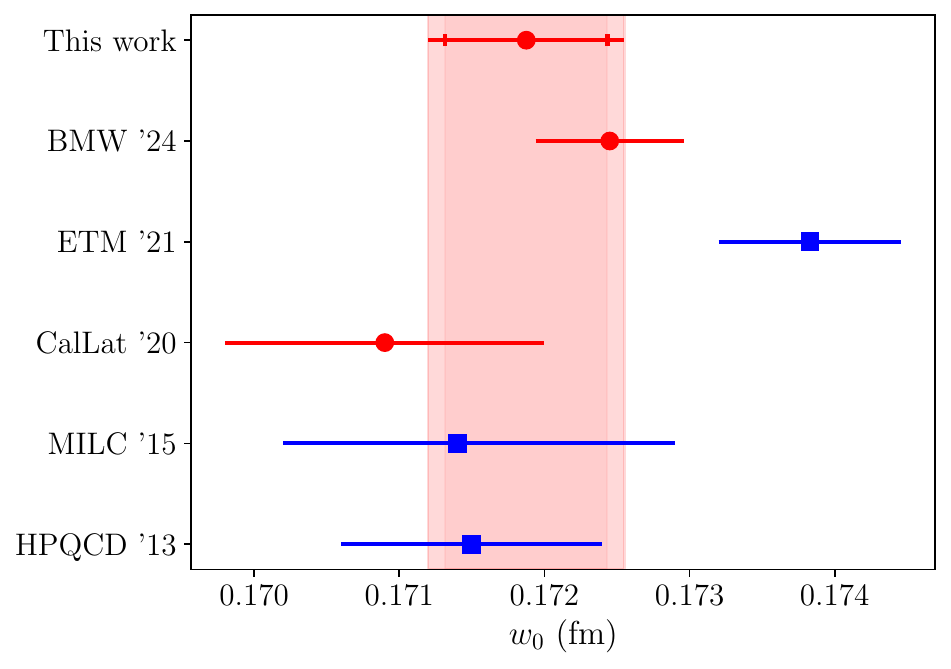}
    \caption{Comparison of the result in this work with those of previous calculations of $w_0$ with $2+1+1$ flavors.
    The inner and outer bands depict statistical and total uncertainties.
    Calculations defined via $M_\Omega$ ($f_\pi$) are shown in red (blue).
    Previous results are BMW~'24~\cite{Boccaletti:2024guq}, ETM~'21~\cite{ExtendedTwistedMass:2021qui}, CalLat~'20~\cite{Miller:2020evg}, MILC~'15~\cite{MILC:2015tqx}, and HPQCD~'13~\cite{Dowdall:2013rya}.}
    \label{fig:result-comparison}
\end{figure}

We compare our result for the gradient-flow scale in Eq.~(\ref{eq:w0-final}) with those of previous calculations in Fig.~\ref{fig:result-comparison}.
This work achieves uncertainties competitive with the most precise determinations by BMW~\cite{Borsanyi:2020mff, Boccaletti:2024guq} and ETM~\cite{ExtendedTwistedMass:2021qui}.
Our final $w_0$ value agrees well with the most recent $(2+1+1)$-flavor results from BMW~\cite{Boccaletti:2024guq} and CalLat~\cite{Miller:2020evg} using the $\Omega^-$ baryon, as well as older results from MILC~\cite{MILC:2015tqx} and HPQCD~\cite{Dowdall:2013rya} using the pion decay constant, $f_\pi$.
(Superseded calculations by BMW~\cite{Borsanyi:2020mff} and ETM~\cite{ExtendedTwistedMass:2020tvp} are also consistent.)
There is some tension, however, between the most recent ETM result~\cite{ExtendedTwistedMass:2021qui}, determined using $f_\pi$ scale setting, and all other scale-setting results, including the one reported here.

En route to Eq.~(\ref{eq:w0-final}), we determined a pure-QCD value of the $\Omega$ baryon mass, consistent with the Edinburgh consensus of FLAG~\cite{FlavourLatticeAveragingGroupFLAG:2024oxs}.
In ongoing work, we will determine the continuum limit of $f_\pi/M_\Omega$ in pure QCD and in QCD+QED.
In future work, we plan to compute the electromagnetic corrections stemming from valence-sea and sea-sea photon exchange, thereby removing the electroquenching uncertainties in~$M_\Omega$.
We will also improve statistics, especially on the finest ($a\approx0.04$~fm) ensemble, to improve the precision of the continuum limit.

\acknowledgments

%%NERSC
This research used resources (especially Franklin and Perlmutter) of the National Energy Research Scientific Computing Center (NERSC), a U.S.\ Department of Energy Office of Science User Facility located at Lawrence Berkeley National Laboratory, operated under Contract No.\ DE-AC02-05CH11231.
%% DOE/INCITE/ALCC
We also used Mira, Theta, and Polaris at the Argonne Leadership Computing Facility (ALCF) and Summit, Crusher, and Frontier at the Oak Ridge Leadership Computing Facility (OLCF) under the Innovative and Novel Computational Impact on Theory and Experiment (INCITE) 
and ASCR Leadership Computing Challenge (ALCC) programs.
The ALCF and OLCF are DOE Office of Science User Facilities supported under contract Nos.\ DE-AC02-06CH11357 and DE-AC05-00OR22725, respectively.
%%USQCD  
Computations for this work were carried out in part with computing and long-term storage resources provided by the USQCD Collaboration. 
%%XSEDE ACCESS   Ranch , Frontera                 
This work used the Extreme Science and Engineering Discovery Environment (XSEDE) storage system Ranch at the Texas Advanced Computing Center (TACC) through allocation TG-MCA93S002. 
The XSEDE program was supported by the National Science Foundation under grant No.\ ACI-1548562.
This work also used Ranch at TACC through allocation MCA93S002 from the Advanced Cyberinfrastructure Coordination Ecosystem: Services \& Support (ACCESS) program, which is supported by U.S.\ National Science Foundation grants Nos.\ 2138259, 2138286, 2138307, 2137603, and 2138296.
Also through ACCESS allocation MCA93S002 this research used both the DeltaAI advanced computing and data resource, which is supported by the National Science Foundation (award OAC 2320345) and the State of Illinois, and the Delta advanced computing and data resource which is supported by the National Science Foundation (award OAC 2005572) and the State of Illinois. Delta and DeltaAI are joint efforts of the University of Illinois Urbana-Champaign and its National Center for Supercomputing Applications.
This research is part of the Frontera computing project at the Texas Advanced Computing Center.
Frontera is made possible by National Science Foundation award OAC-1818253.
%%Big Red II+ , etc.                                       
Computations on the Big Red II+, Big Red 3, Quartz, and Big Red 200 computers were supported in part by Lilly Endowment, Inc., through its support for the Indiana University Pervasive Technology Institute.
The parallel file system employed by Big Red II+ was supported by the National Science Foundation under Grant No.~CNS-0521433.

%%Blue Waters                               
Some of the computations were done using the Blue Waters sustained-petascale computer, which was supported by the National Science Foundation (awards OCI-0725070 and ACI-1238993) and the state of Illinois.
Blue Waters was a joint effort of the University of Illinois at Urbana-Champaign and its National Center for Supercomputing Applications (NCSA).

This work was in part based on the MILC collaboration's public lattice gauge theory code~\cite{milc} with QUDA~\cite{Clark:2009wm, Babich:2011np, QUDAgithub} used to accelerate quark propagator solves and smearing on~GPUs.

This work was supported in part by the U.S.~Department of Energy, Office of Science, under Awards
No.~DE-SC0010005 (E.T.N. and J.W.S.),
No.~DE-SC0010120 (S.G.), 
No.~DE-SC0015655 (A.X.K., S.L., M.L., A.T.L.),
No.~DE-SC0009998 (J.L.),
No.~DE-SC0011090 (Y.L.),
the Neutrino Theory Network Program Grant No.\ DE-AC02-07CHI11359 and No.\ DE-SC0020250 (A.S.M.),
the ``High Energy Physics Computing Traineeship for Lattice Gauge Theory'' No.~DE-SC0024053 (J.W.S.),
and the Funding Opportunity Announcement Scientific Discovery through Advanced Computing: High Energy Physics, LAB 22-2580 (D.A.C., L.H., M.L., S.L., C.T.P.);
by the Exascale Computing Project (17-SC-20-SC), a collaborative effort of the U.S. Department of Energy Office of Science and the National Nuclear Security Administration (H.J.);
by the National Science Foundation under Grants Nos.~PHY20-13064 and PHY23-10571 (C.D., D.A.C., S.L., A.V.),
No.~PHY23-09946 (A.B.),
No.~PHY-2402275 (A.V.G.), and
No.~2139536 for Characteristic Science Applications for the Leadership Class Computing Facility (L.H., H.J.);
by the Simons Foundation under their Simons Fellows in Theoretical Physics program (A.X.K.); 
by the Universities Research Association Visiting Scholarship awards 20-S-12 and 21-S-05 (S.L.);  
by MICIU/AEI/10.13039/501100011033 and FEDER (EU) under Grant PID2022-140440NB-C21 (E.G.);
by Consejeria de Universidad, Investigaci\'on y Innovaci\'on and Gobierno de Espa\~na and EU--NextGenerationEU, under Grant AST22~8.4 (E.G.);
by AEI (Spain) under Grant No.\ RYC2020-030244-I / AEI / 10.13039/501100011033 (A.V.);
%%% visits, workshops, programs
A.X.K.\ and E.T.N.\ are grateful to the Pauli Center for Theoretical Studies and the ETH Z\"urich for support and hospitality.
A.X.K., A.S.K., and E.T.N.\ are grateful to the Kavli Institute for Theoretical Physics (KITP) for hospitality and support during the program ``What is Particle Theory?''
The KITP is supported in part by the National Science Foundation under Grant No.\ PHY-2309135.
A.B.\ and A.V.G.\ thank ECT* for support at the Workshop ``Scale setting: Precision lattice QCD for particle and nuclear physics'' during which part of this work was developed.
This work was performed under the auspices of the U.S.\ Department of Energy by Lawrence Livermore National Laboratory under Contract No.\ DE-AC52-07NA27344 (A.S.M.).
This document was prepared by the Fermilab Lattice and MILC Collaborations using the resources of the Fermi National Accelerator Laboratory (Fermilab), a U.S.\ Department of Energy, Office of Science, HEP User Facility.
Fermilab is managed by Fermi Forward Discovery Group, LLC, acting under Contract No.~89243024CSC000002 with the U.S.\ Department of Energy.

\appendix

\section{Bayesian model averaging}
\label{app:BMA}

Various correlator fits are combined using the Bayesian model-averaging (BMA) procedure introduced in Refs.~\cite{BMW:2014pzb,Berkowitz:2017gql,Chang:2018uxx,Jay:2020jkz,Neil:2022joj}.
Specifically, for a given model $i$, one defines  the Bayesian Akaike information criterion (BAIC), a modified Akaike information criterion to account for data excluded from the fit range~\cite{Neil:2022joj,Neil:2023pgt}, as
\begin{equation}
    \text{BAIC}_i = \chi_{\rm data}^{2}\left(\mathbf{a}^{\star}\right)+2 k+2 N_{\mathrm{cut}}
    \label{eq:aic}
\end{equation}
where $\chi^2_\text{data}(\mathbf{a}^\star)$ is the contribution to the $\chi^2$ function from the data (as opposed to the priors) evaluated at the minimum $\mathbf{a}^\star$ of the augmented $\chi^2$ function (including priors)~\cite{Lepage:2001ym}, $k$ is the number of fit parameters, and $N_\text{cut}$ is the number of data points removed from the fit range.  The weight of the model is defined as its probability given the data, computed as
\begin{equation}
    w_i \propto \exp\left[ - \frac{1}{2} \text{BAIC}_i \right]
    \label{eq:model-weights}
\end{equation}
normalized such that $\sum_i w_i = 1$.  The model average of the ground-state energies $E_i$ measured on the various models is given by
\begin{equation}
    \langle E \rangle = \sum_i w_i E_i
    \label{eq:bma}
\end{equation}
with statistical and systematic uncertainties given by
\begin{align}
    (\sigma_E^2)_\text{stat} &= \sum_i w_i (\sigma_E^2)_i \, , \label{eq:bma-stat-error} \\
    (\sigma_E^2)_\text{sys} &= \sum_i w_i \left(E_i - \langle E \rangle\right)^2 \, , \label{eq:bma-sys-error}
\end{align}
which are added in quadrature to give the full uncertainty~$\sigma_E^2$.

\section{Error amplification from quark-mass corrections}
\label{app:stuff}

While $\gamma_v$ and $\gamma_s$ are nuisance parameters from the standpoint of scale setting, they are physical quantities that encode the variation of the $\Omega$ baryon mass and $w_0$ with quark masses.
Performing a BMA over fits gives
\begin{align}
    \gamma_v^{(0)} &= 0.55(5) , \\
    \gamma_s^{(0)} &= 0.08(3) .
\end{align}
The error bars contain both statistical and systematic uncertainties, and the values are of magnitudes similar to those computed from single-ensemble fits in Ref.~\cite{Ren:2013oaa}.
The strange-quark mass dependence $d \ln M_\Omega / d \ln m_s \approx \gamma_v / (1 + 2\gamma_v + 2\gamma_s) = 0.24(1)$ can be computed from these quantities.
Recall that this logarithmic derivative is needed to obtain the mass-matching factor $\zeta$ between QCD+QED and QCD.
In practice, a few iterations of the whole procedure are needed until it stabilizes.

This substantial dependence of $M_\Omega$ on quark masses increases the uncertainty in quark-mass-corrected values of $M_\Omega$.
Suppose that $aM_\Omega$ has been determined on a given ensemble with relative uncertainty $\delta M_\Omega$.
Neglecting the (much smaller) uncertainties in meson masses, this induces a relative uncertainty of
\begin{equation}
    \delta v_\Omega = \delta s_\Omega = -2 \delta M_\Omega
\end{equation}
in the proxies used for valence and sea quark masses.
Thus, the quark-mass-corrected $\Omega$ baryon mass $M_\Omega^\text{corr}$ has an uncertainty given by
\begin{align}
    \delta M_\Omega^\text{corr} &= \delta M_\Omega - \gamma_v \delta v_\Omega - \gamma_s \delta s_\Omega \\
        &= (1 + 2\gamma_v + 2\gamma_s)\delta M_\Omega ,
\end{align}
that is, the original uncertainty has been amplified by the factor $1 + 2\gamma_v + 2\gamma_s = 2.2(1)$, leading to the larger error bars in the quark-mass-corrected values in the left panel of Fig.~\ref{fig:continuum-extrapolation-multiple}.

\begin{figure}
  \centering
  \includegraphics[width=0.5\textwidth]{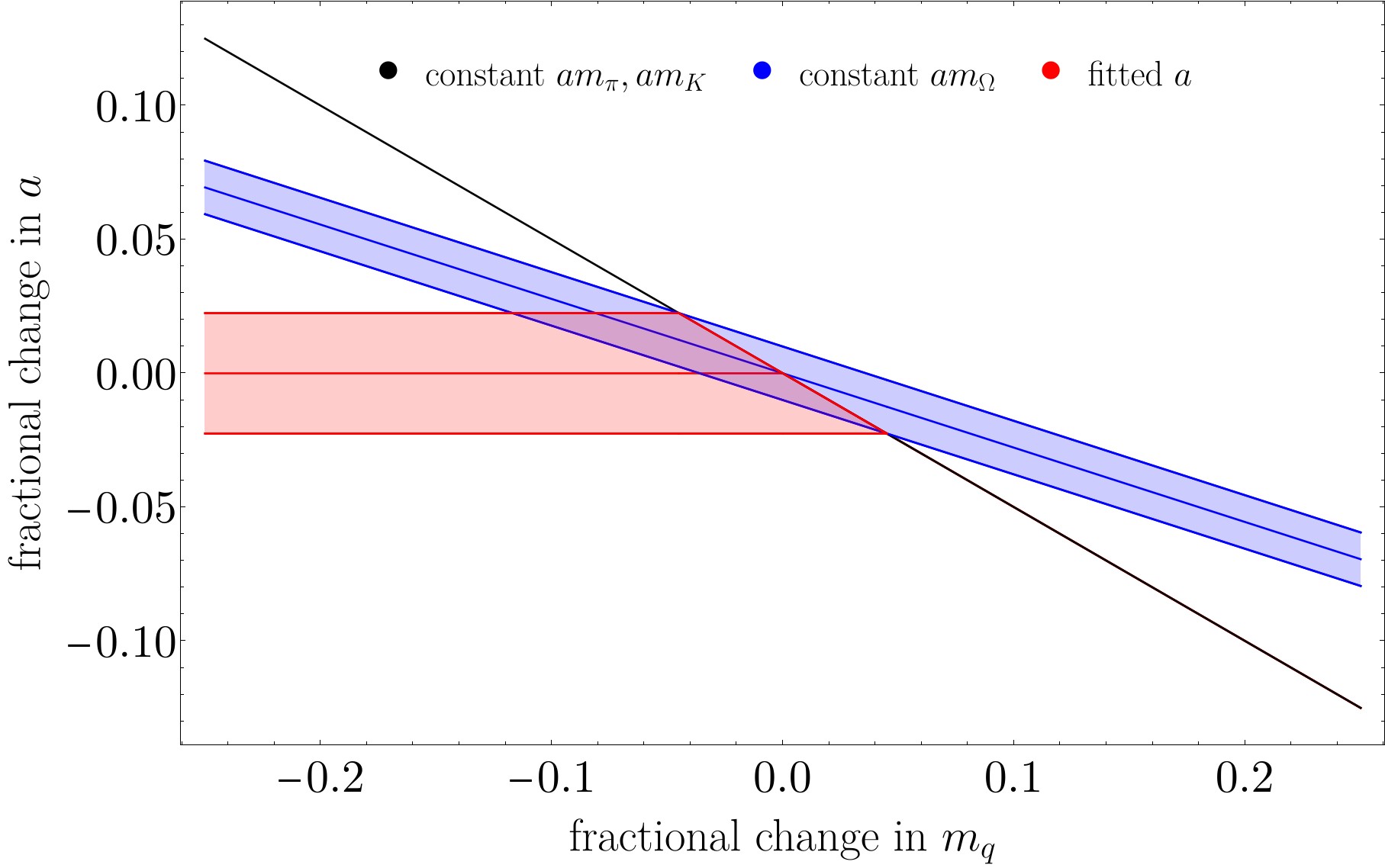}
  \caption{An illustration of the error inflation inherent in self-consistently solving for the lattice spacing and quark masses using the $\Omega$ baryon and pseudoscalar meson masses.  The black and blue lines represent contours of constant $aM_\text{PS}$ and $aM_\Omega$, respectively, which have slopes $-1/2$ and $-(\gamma_v+\gamma_s)/(1+2\gamma_v+2\gamma_s)\approx -0.28$ from leading-order $\chi$PT when plotted in the $(\log m_q, \log a)$ plane.  Assuming that the mesons can be measured precisely, the uncertainty in the quantity of interest ($a$) comes entirely from that in $M_\Omega$, where the blue band represents a 1\% relative uncertainty in the measurement.  Propagating this uncertainty through the self-consistent solution of equations to extract $a$ inflates this uncertainty by a factor $1+2\gamma_v+2\gamma_s \approx 2.2$ indicated by the red band.}
  \label{fig:error-inflation-plot}
\end{figure}

This error-inflation factor can be thought of as the Jacobian of the change-of-basis between the measured quantities $\left\{ aM_\Omega, (a m_\text{val})^2, (a m_\text{sea})^2 \right\}$ to the desired results $\left\{ a, m_\text{val}, m_\text{sea} \right\}$.  In particular, this factor is present even if $\gamma_v$ and $\gamma_s$ are known exactly and even if the ensembles are well enough tuned that the quark mass corrections are consistent with zero.  This uncertainty inflation is illustrated graphically in Fig.~\ref{fig:error-inflation-plot}.

\section{Results for \texorpdfstring{\boldmath$aM_\Omega$}{aMOmega} and \texorpdfstring{\boldmath$w_0/a$}{w0/a}}
\label{app:lattice-results}

This Appendix tabulates results for the $\Omega$-baryon mass and gradient-flow scale on each ensemble.
Results for $aM_\Omega$ from the correlator fits of Sec.~\ref{sec:corrfits} are given in Table~\ref{tab:omega-masses}, that is, without strange-quark-mass retuning.
To aid the retuning, meson masses (in lattice units) are provided in Table~\ref{tab:meson-masses}.
Results for the gradient flow $w_0/a$ for each of the twelve definitions, as described in Sec.~\ref{sec:gradient-flow}, are listed in Table~\ref{tab:w0-scale}.

\begin{table}
    \centering
    \caption{Results for $aM_\Omega$ from the correlator fits of Sec.~\ref{sec:corrfits} for the three tastes of the $\Omega$ baryon.
    For ensembles with two valence masses, the result for the unitary valence mass is given first, followed by that for the heavier-than-unitary valence mass, and the correlation coefficient between the two results is given in brackets.}
    \label{tab:omega-masses}
    \begin{tabular}{ccccccc}
        \hline\hline
        $\approx a$/fm & & $aM_\Omega^\text{mixed}$ & $aM_\Omega^{44}$ & $aM_\Omega^{77}$ \\
         \hline
        0.15 & $\left\{ \begin{matrix} \\ \\ \\ \end{matrix} \right.$ & $\begin{matrix} 1.30102(88) \\ 1.30374(87) \\ [0.999938] \end{matrix}$ & $\begin{matrix} 1.30017(97) \\ 1.30289(96) \\ [0.999946] \end{matrix}$ & $\begin{matrix} 1.29991(90) \\ 1.30264(89) \\ [0.999918] \end{matrix}$ \\
        0.12H & & 1.0366(14) & 1.0352(17) & 1.0357(13) \\
        0.12 & $\left\{ \begin{matrix} \\ \\ \\ \end{matrix} \right.$ & $\begin{matrix} 1.03745(65) \\ 1.03971(63) \\ [0.999219] \end{matrix}$ & $\begin{matrix} 1.03718(46) \\ 1.03946(45) \\ [0.999527] \end{matrix}$ & $\begin{matrix} 1.03717(62) \\ 1.03945(62) \\ [0.999367] \end{matrix}$ \\
        0.09H & & 0.75443(82) & 0.75382(93) & 0.75421(84) \\
        0.09M & $\left\{ \begin{matrix} \\ \\ \\ \end{matrix} \right.$ & $\begin{matrix} 0.74613(60) \\ 0.74641(61) \\ [0.999917] \end{matrix}$ & $\begin{matrix} 0.74598(49) \\ 0.74626(49) \\ [0.999646] \end{matrix}$ & $\begin{matrix} 0.74647(40) \\ 0.74675(40) \\ [0.999952] \end{matrix}$ \\
        0.09C & $\left\{ \begin{matrix} \\ \\ \\ \end{matrix} \right.$ & $\begin{matrix} 0.74851(37) \\ 0.75023(36) \\ [0.998037] \end{matrix}$ & $\begin{matrix} 0.74844(34) \\ 0.75014(34) \\ [0.999798] \end{matrix}$ & $\begin{matrix} 0.74873(40) \\ 0.75042(39) \\ [0.999240] \end{matrix}$ \\
        0.09L & & 0.74798(19) & 0.74753(28) & 0.74766(14) \\
        0.06 & & 0.48257(48) & 0.48249(70) & 0.48266(70) \\
        0.04 & & 0.3599(11) & 0.36040(64) & 0.36014(96) \\
        \hline\hline
    \end{tabular}
\end{table}

\begin{table}
    \centering
    \caption{Results for the sea-meson masses used for quark-mass tuning.
    Values on the physical-point ensembles (0.15, 0.12, 0.09M, 0.09C, 0.06, and 0.04~fm) are taken from Ref.~\cite{MILC:2024ryz} and references therein.
    Values on the heavier-than-physical ensembles 0.12H and 0.09H are taken from Refs.~\cite{FermilabLattice:2022gku} and~\cite{Bazavov:2017lyh}, respectively.
    The value on the large-volume ensemble 0.09L is new.
    In all cases, values are determined from multi-exponential fits to the truncated spectral decomposition using methods and parameters described in Refs.~\cite{Bazavov:2017lyh,FermilabLattice:2022gku}.}
    \label{tab:meson-masses}
    \begin{tabular}{ccc}
        \hline\hline
        $\approx a$/fm & $aM_\pi$ & $aM_K$ \\ \hline
        0.15 & 0.103414(11) & 0.37847(11) \\
        0.12H & 0.13426(8) & 0.30806(12) \\
        0.12 & 0.0830651(63) & 0.303949(77) \\
        0.09H & 0.09860(13) & 0.22706(15) \\
        0.09M & 0.057184(30) & 0.219482(70) \\
        0.09C & 0.060069(26) & 0.220231(40) \\
        0.09L & 0.0600523(72) & 0.220225(13) \\
        0.06 & 0.038842(29) & 0.142607(51) \\
        0.04 & 0.028981(18) & 0.106297(39) \\
        \hline\hline
    \end{tabular}
\end{table}

\begin{table*}
    \centering
    \caption{Results for the gradient-flow scale $w_0/a$ for each of the twelve combinations of flow, improvement, and operator.}
    \label{tab:w0-scale}
    \begin{tabular}{cccccc}
        \hline\hline
         $\approx a$/fm & Flow & Correction & Clover & Wilson & Symanzik \\ \hline
         \multirow{4}{*}{0.15} & \multirow{2}{*}{Symanzik} & no & 1.14272(18) & 1.13067(18) & 1.12784(19) \\
         & & yes & 1.13128(20) & 1.12168(20) & 1.11795(19) \\
         & \multirow{2}{*}{Wilson} & no & 1.13227(18) & 1.15111(18) & 1.15825(19) \\
         & & yes & 1.13359(21) & 1.14303(19) & 1.14421(19) \\ \hline
         
         \multirow{4}{*}{0.12H} & \multirow{2}{*}{Symanzik} & no & 1.40407(82) & 1.39489(85) & 1.39251(86) \\
         & & yes & 1.40891(87) & 1.39402(87) & 1.38916(87) \\
         & \multirow{2}{*}{Wilson} & no & 1.40302(83) & 1.40936(86) & 1.41230(86) \\
         & & yes & 1.41301(87) & 1.40686(87) & 1.40466(87) \\ \hline

         \multirow{4}{*}{0.12} & \multirow{2}{*}{Symanzik} & no & 1.41156(27) & 1.40262(28) & 1.40033(28) \\
         & & yes & 1.41701(29) & 1.40198(29) & 1.39707(29) \\
         & \multirow{2}{*}{Wilson} & no & 1.41060(28) & 1.41729(29) & 1.42035(29) \\
         & & yes & 1.42110(30) & 1.41489(29) & 1.41267(29) \\ \hline

         \multirow{4}{*}{0.09H} & \multirow{2}{*}{Symanzik} & no & 1.92854(130) & 1.92405(132) & 1.92285(133) \\
         & & yes & 1.93917(134) & 1.92691(134) & 1.92291(133) \\
         & \multirow{2}{*}{Wilson} & no & 1.93328(131) & 1.93522(134) & 1.93622(135) \\
         & & yes & 1.94453(135) & 1.93568(135) & 1.93279(134) \\ \hline
         
         \multirow{4}{*}{0.09M} & \multirow{2}{*}{Symanzik} & no & 1.94664(40) & 1.94251(40) & 1.94144(41) \\
         & & yes & 1.95796(42) & 1.94563(41) & 1.94161(41) \\
         & \multirow{2}{*}{Wilson} & no & 1.95148(41) & 1.95386(42) & 1.95501(42) \\
         & & yes & 1.96328(42) & 1.95444(42) & 1.95155(42) \\ \hline
         
         \multirow{4}{*}{0.09C} & \multirow{2}{*}{Symanzik} & no & 1.94537(56) & 1.94121(57) & 1.94012(58) \\
         & & yes & 1.95663(58) & 1.94430(58) & 1.94028(58) \\
         & \multirow{2}{*}{Wilson} & no & 1.95021(57) & 1.95255(58) & 1.95368(58) \\
         & & yes & 1.96195(59) & 1.95312(58) & 1.95023(58) \\ \hline
         
         \multirow{4}{*}{0.09L} & \multirow{2}{*}{Symanzik} & no & 1.94506(30) & 1.94091(31) & 1.93982(31) \\
         & & yes & 1.95632(31) & 1.94400(31) & 1.93999(31) \\
         & \multirow{2}{*}{Wilson} & no & 1.94989(31) & 1.95223(32) & 1.95336(32) \\
         & & yes & 1.96163(32) & 1.95280(31) & 1.94991(32) \\ \hline
         
         \multirow{4}{*}{0.06} & \multirow{2}{*}{Symanzik} & no & 3.01269(91) & 3.01215(92) & 3.01207(92) \\
         & & yes & 3.02358(92) & 3.01587(92) & 3.01333(92) \\
         & \multirow{2}{*}{Wilson} & no & 3.01838(92) & 3.01978(92) & 3.02034(92) \\
         & & yes & 3.02786(93) & 3.02116(92) & 3.01896(92) \\ \hline
         
         \multirow{4}{*}{0.04} & \multirow{2}{*}{Symanzik} & no & 4.02741(194) & 4.02772(194) & 4.02786(194) \\
         & & yes & 4.03663(195) & 4.03099(195) & 4.02912(195) \\
         & \multirow{2}{*}{Wilson} & no & 4.03242(195) & 4.03358(196) & 4.03402(196) \\
         & & yes & 4.04009(196) & 4.03490(196) & 4.03318(196) \\
         \hline\hline
    \end{tabular}
\end{table*}

\section{Continuum-Extrapolated \texorpdfstring{\boldmath$w_0 M_\Omega$}{w0*MOmega} Schemes}
\label{app:12w0MOmega}

While different definitions of $w_0/a$ (characterized by different choices of the flow and observable) disagree at finite lattice spacing, they should all provide the same continuum limit for $w_0 M_\Omega$.
Nevertheless, to facilitate comparisons with previous work using a specific definition of $w_0/a$, we tabulate in Table~\ref{tab:w0-scale-continuum} values using the twelve different definitions.
They are shown in Fig.~\ref{fig:continuum-extrapolation-summary} and their Bayesian model average gives the overall result in the main text.

\begin{table}
    \centering
    \caption{Continuum extrapolations of $w_0 M_\Omega$ for the twelve combinations of flow, improvement, and operator described in Sec.~\ref{sec:gradient-flow}.}
    \label{tab:w0-scale-continuum}
    \begin{tabular}{ccccc}
        \hline\hline
         Flow & Correction & Clover & Wilson & Symanzik \\
        \hline
         \multirow{2}{*}{Symanzik} & no & 1.4536(60) & 1.4552(61) & 1.4557(61) \\
             & yes & 1.4555(55) & 1.4555(43) & 1.4556(51) \\
         \multirow{2}{*}{Wilson} & no & 1.4552(44) & 1.4559(70) & 1.4554(76) \\
             & yes & 1.4558(60) & 1.4566(61) & 1.4563(68) \\
         \hline\hline
    \end{tabular}
\end{table}

\clearpage
\bibliographystyle{apsrev4-2}
\bibliography{bib}

\end{document}